\documentclass[english,aps,reprint,nofootinbib,longbibliography,notitlepage,superscriptaddress,prd,preprintnumbers,showkeys]{revtex4-1}
\usepackage[latin9]{inputenc}
\usepackage{babel}
\usepackage{float}
\usepackage{graphicx}
\usepackage{hyperref}
\usepackage{amsmath}
\usepackage{amssymb}
\usepackage{todonotes}
\usepackage[mathscr]{eucal}
\usepackage{microtype}
\usepackage{esint}
\usepackage{amsbsy}
\usepackage{color}
\usepackage{cleveref}
\usepackage{breakurl}
\usepackage{ulem}
\usepackage{bbm}
\usepackage{mathtools}
\usepackage{braket}
\usepackage{comment}
\usepackage[super]{nth}
\usepackage{todonotes}
\usepackage{hyphenat}

\makeatletter

\pdfpageheight\paperheight
\pdfpagewidth\paperwidth

\@ifundefined{textcolor}{}{%
 \definecolor{BLACK}{gray}{0}
 \definecolor{WHITE}{gray}{1}
 \definecolor{RED}{rgb}{1,0,0}
 \definecolor{GREEN}{rgb}{0,1,0}
 \definecolor{BLUE}{rgb}{0,0,1}
 \definecolor{CYAN}{cmyk}{1,0,0,0}
 \definecolor{MAGENTA}{cmyk}{0,1,0,0}
 \definecolor{YELLOW}{cmyk}{0,0,1,0}
}

\makeatother

\makeatletter

\DeclareRobustCommand{\rcite}[1]{%
  \rcite@aux#1,\@nil{#1}%
}
\def\rcite@aux#1,#2\@nil#3{%
  \if\relax#2\relax
    Ref.~\cite{#3}%
  \else
    Refs.~\cite{#3}%
  \fi
}
\makeatother

\hypersetup{
    colorlinks = true,
    citecolor = {red},
    linkcolor = {blue},
    urlcolor = {blue},
}

\newcommand{\be}{\begin{equation}}
\newcommand{\ee}{\end{equation}}
\newcommand{\lagr}{{\cal L}}

\DeclareMathOperator\cd{cd}

\begin{document}

\title{The Classical Symmetron Force in Casimir Experiments}

\author{Benjamin Elder}
\email{benjamin.elder@nottingham.ac.uk}
\affiliation{School of Physics and Astronomy, University of Nottingham, Nottingham, NG7 2RD, United Kingdom}

\author{Valeri Vardanyan}
\email{vardanyan@lorentz.leidenuniv.nl}
\affiliation{Lorentz Institute for Theoretical Physics, Leiden University, P.O. Box 9506, 2300 RA Leiden, The Netherlands}
\affiliation{Leiden Observatory, Leiden University, P.O. Box 9513, 2300 RA Leiden, The Netherlands}
\affiliation{Kavli Institute for the Physics and Mathematics of the Universe (WPI), UTIAS, The University of Tokyo, Chiba 277-8583, Japan}

\author{Yashar Akrami}
\email{akrami@ens.fr}
\affiliation{Laboratoire de Physique de l'\'Ecole Normale Sup\'erieure, ENS, Universit\'e PSL, CNRS, Sorbonne Universit\'e, Universit\'e de Paris, F-75005 Paris, France}
\affiliation{Observatoire de Paris, Universit\'e PSL, Sorbonne Universit\'e, LERMA, 75014 Paris, France}

\author{Philippe Brax}
\email{philippe.brax@ipht.fr}
\affiliation{Institut de Physique Th{\'e}orique, Universit{\'e} Paris-Saclay, CEA, CNRS, F-91191 Gif/Yvette Cedex, France}

\author{Anne-Christine Davis}
\email{acd@damtp.cam.ac.uk}
\affiliation{DAMTP, Centre for Mathematical Sciences, University of Cambridge, CB3 0WA, United Kingdom}

\author{Ricardo S. Decca}
\email{rdecca@iupui.edu}
\affiliation{Department of Physics, Indiana University-Purdue University Indianapolis, 402 N. Blackford St., Bldg LD154, Indianapolis, IN 46202, USA}


\begin{abstract}
The symmetron is a typical example of screened modified gravity, wherein the symmetron force is dynamically suppressed in dense environments.  This allows it to hide in traditional tests of gravity. However, the past decade has seen great experimental progress towards measuring screened forces in the laboratory or in space. Screening relies on nonlinearities in the equation of motion, which significantly complicates the theoretical analysis of such forces.  Here, we present a calculation of the symmetron force between a dense plate and sphere surrounded by vacuum.  This is done via semi-analytical approaches in two limiting cases, based on the size of the sphere: large spheres are analyzed via the proximity force approximation, whilst small spheres are treated as screened test particles.  In the intermediate regime we solve the problem numerically.  Our results allow us to make contact with Casimir force experiments, which often employ a plate and sphere configuration for practical reasons, and may therefore be used to constrain symmetrons.  We use our results to forecast constraints on the symmetron's parameters for a hypothetical Casimir experiment that is based on the current state of the art.  The forecasts compare favorably to other leading laboratory tests of gravity, particularly atom interferometry and bouncing neutrons. We thus conclude that near-future Casimir experiments will be capable of placing tight new bounds on symmetrons.  Our results for the symmetron force are derived in a scale-invariant way, such that although we here focus on Casimir experiments, they may be applied to any other plate-sphere system, ranging from microscopic to astrophysical scales.
\end{abstract}

\maketitle

\tableofcontents

\section{Introduction}
\label{sec:intro}

The accelerated expansion of the Universe has motivated new  classes of scalar-tensor theories~\cite{Khoury:2010xi,Brax:2017idh} that go beyond general relativity (GR)~\cite{Einstein:1916vd}. As the coupling of the scalar sector to matter is generically  not explicitly forbidden by the symmetries of these  scalar-tensor theories, there is in general no theoretical restriction on the existence of  scalar ``fifth forces'' between matter particles. On the other hand, such forces are experimentally tightly constrained in the Solar System and by laboratory experiments~\cite{Will:1993ns,Burrage:2017qrf,Brax:2018iyo}. Compatibility with these bounds then requires the fifth force to be either tuned to be extremely weak, or to be dynamically suppressed in certain situations.

This latter approach, dubbed as ``screening'', has generated a great deal of interest over the past two decades~\cite{Brax:2013ida}. There are four main classes of screening mechanisms that have been discovered so far (see \cite{Joyce:2014kja} for a comprehensive review):
\begin{enumerate}
\item \textit{\textbf{chameleons~\cite{Khoury:2003rn}:}} The scalar field becomes heavy, and therefore short-ranged, in dense environments.
\item \textit{\textbf{Damour-Polyakov~\cite{Damour:1994zq}:}} The scalar field  decouples  when the coupling to matter is dynamically driven to zero. Examples are given by the symmetron~\cite{Pietroni:2005pv,Olive:2007aj,Hinterbichler:2010es, Hinterbichler:2011ca} and the dilaton in the large string coupling limit~\cite{Brax:2010gi}.
\item \textit{\textbf{K-mouflage}:} The steep field gradients of the scalar field help to suppress fifth forces~\cite{Babichev:2009ee,Brax:2012jr,Brax:2014yla}.
\item \textit{\textbf{Vainshtein}:} Fifth forces are suppressed in regions where the second derivatives of the scalar field are sufficiently large~\cite{Vainshtein:1972sx, Nicolis:2008in}.
\end{enumerate}
The phenomenology of the first two mechanisms turns out to be similar, i.e. they both screen fifth forces where the surface Newtonian potential of the body is sufficiently large~\cite{Brax:2012gr}.
For concreteness, in this paper we focus on symmetrons, even though extension to other scenarios, particularly chameleons, is possible.

Symmetron gravity has been extensively studied in contexts ranging from cosmological large-scale structure to laboratory experiments. As mentioned above, the main observable effect of symmetrons is that matter feels an extra force compared to the general relativistic gravitational interaction. Hence the large-scale dark matter distribution in the Universe is affected and differs from the GR case. Such changes are investigated in Refs.~\cite{Davis:2011pj, Brax:2012nk}. On smaller scales, symmetrons are naturally expected to alter the collapse dynamics of dark matter shells, and, as such, leave  observable imprints on the density profiles of dark matter halos~\cite{Contigiani:2018hbn}, as well as to alter the halo mass function~\cite{Taddei:2013bsk}. Moreover, symmetron fifth forces will necessarily alter the velocity dispersion of dark matter tracers in halos, therefore leading to a potential mismatch between the dynamical halo mass and the lensing mass. This feature is exploited as a test of symmetron gravity in Ref.~\cite{Clampitt:2011mx}.
Another interesting class of tests is suggested at galactic and stellar scales. Particularly, the effects on stellar dynamics in galaxies and globular clusters are explored in Refs.~\cite{OHare:2018ayv, Llinares:2018dtu}. Refs.~\cite{Brax:2013uh, Zhang:2017srh} consider changes in energy output from binary pulsars, Ref.~\cite{Davis:2011qf} investigates the effect of screening on stellar evolution and Ref.~\cite{Jain:2012tn} studies changes in stellar physics induced by the symmetron force. The presence of a scalar force is also expected to induce a mismatch between the stellar and gas components in galaxies. This idea in the context of symmetron gravity is explored in Ref.~\cite{Desmond:2018euk}.

Besides cosmological and astrophysical tests, symmetrons can also be tested in laboratory experiments. Notably, {\it atom interferometry} experiments have been able to constrain them significantly~\cite{Burrage:2016rkv, Jaffe:2016fsh, Sabulsky:2018jma}. These are complemented by {\it torsion balance} experiments~\cite{Upadhye:2012rc,Brax:2014zta}. Constraints also follow from quantum effects of the symmetron which affect the anomalous magnetic moment of the electron~\cite{Brax:2018zfb} and lead to corrections to the energy levels of neutrons in the gravitational field of the Earth~\cite{Cronenberg:2018qxf}.

In this paper, we focus on new bounds on symmetrons that may be obtained from Casimir force sensors~\cite{Lamoreaux:1996wh,Decca:2005yk,Decca:2003zz}. Although conceived as detectors for the quantum mechanical Casimir-Polder force, these systems are  also sensitive to any additional forces~\cite{Decca:2005qz,Decca:2003zz}, including classical screened fifth forces~\cite{Brax:2007vm}.  The advantage of using  Casimir force sensors to look for new forces is that they probe very short distance scales ($\sim 10$ $\mu$m) and therefore larger particle masses compared to longer-range fifth force detectors in the laboratory or the Solar System (see, for instance, Ref.~\cite{Berge:2017ovy}).  Casimir experiments have been used to constrain chameleons~\cite{Brax:2007vm}, and new experiments~\cite{Brax:2010xx} will provide results soon~\cite{Almasi:2015zpa,Sedmik:2018kqt}.  However, such constraints have not yet been presented for symmetrons, mainly due to a lack of theoretical calculations of the symmetron force between extended objects~\cite{Burrage:2017qrf}.

In this article we address this shortcoming by computing the symmetron force between a plate and a sphere.  We focus on this case for two reasons. First, it is perhaps the simplest asymmetric  geometry with two extended objects that one could consider.  Second, our results will be directly applicable to realistic Casimir force sensors, which often use a plate and a sphere for practical experimental reasons~\cite{Chen:2014oda}, rather than two parallel plates (as in some previous experiments such as CANNEX~\cite{Almasi:2015zpa}). This configuration leads to significant theoretical challenges, as screened fifth forces generically have nonlinear equations of motion which are difficult or impossible to solve in less symmetric setups.  In this paper we overcome this hurdle by identifying two limiting regimes that allow  the symmetron force between a sphere and a plate to be approximated analytically.  We also present numerical solutions that interpolate between the limiting cases, and which verify our analytic treatments.

We do this by splitting our analysis into three regions, determined by the size of the dimensionless quantity $\mu R$, where $R$ is the radius of the sphere and $\mu$ is approximately the mass of the symmetron in vacuum.  Physically, this measures the radius of the sphere in units of the symmetron's Compton wavelength.  Our three regions are thus $\mu R \ll 1, \sim 1, $ and $\gg 1$.  In the first and third regions we are able to exploit the hierarchy of scales between the sphere radius and the symmetron Compton wavelength to obtain accurate analytic approximations to  the force.  In the middle region we solve the problem numerically.  It is fortunate that the numerical solution is easiest to obtain when all scales in the problem are comparable, given that this is precisely the regime in which analytic approximations break down. Our results cover several orders of magnitude in sphere radii and separations, and our methodology can be applied to more extreme configurations, not directly reported in this paper. Hence we present a complete solution for the  symmetron force in the plate-sphere system. Furthermore, the solutions we obtain  are fully nonlinear; at no point is the theory linearized about the background vacuum expectation value (apart from the derivation of the screening factor of a dense sphere, which is standard). We then apply our solutions to a realistic Casimir sensor, and show that the results from future measurements will be able to constrain a large part of the symmetron's parameter space.  Although our results are applied to a specific type of problem, i.e. the force between a sphere and a plate which are of laboratory sizes, we emphasize that these solutions are scale-independent and could apply to many other setups such as astrophysical problems.

In this paper we concentrate on the classical symmetron interaction between a sphere and a plate. We do not take into account quantum effects which can be significant. Such quantum effects have been studied in Ref.~\cite{Brax:2018grq} for the plate-plate case, and as a rule, the effects dominate when the scalar field between interacting bodies can be considered almost massless and the self-interaction of the scalar field is large. The forecasts we present here are in the small self-interaction regime, where we expect the quantum effects to be negligible. We leave a comprehensive analysis of the quantum interactions in the plate-sphere system for future work.

This paper is organized as follows. We first review the symmetron screening mechanism in Sec.~\ref{sec:symmetron}. In Sec.~\ref{sec:parallel_plates}, we focus on the two-parallel-plate case. This provides us with intuition for how analytic approximations should be made for the plate-sphere configuration, and additionally, allows us to validate our numerical setup. In Sec.~\ref{sec:analytic}, we analyze the plate-sphere system analytically, and provide analytic approximations for the two cases of large and small spheres. In Sec.~\ref{sec:numerical}, we discuss our procedure for numerically integrating the symmetron equation of motion and computing the force between the sphere and the plate in regimes that are not captured by our analytic results. We show that in the regions of the parameter space where we expect both analytic and numerical approaches to be valid, the results are in very good agreement. We test our numerical framework for a range of configuration parameters and show that it can be applied to any experimental setup. In Sec.~\ref{sec:forecast}, we propose a Casimir force experiment, based on a plate-sphere configuration and with specifications and sensitivity that closely follow those of existing state-of-the-art experiments. We place forecast constraints on the symmetron's free parameters using the expected Casimir bounds on fifth forces. We compare our constraints to those obtained from other laboratory experiments based on atom interferometry~\cite{Jaffe:2016fsh}, torsion balances~\cite{Upadhye:2012rc}, and ultracold bouncing neutron measurements~\cite{Cronenberg:2018qxf}.  Finally, we conclude in Sec.~\ref{sec:conclusions}.

\section{The symmetron model}
\label{sec:symmetron}
\subsection{The model}

The symmetron is a real scalar field with a Lagrangian density\footnote{We are using units where $c = \hbar = 1$ and are working with the mostly-plus metric convention.} given by
\be
\lagr = - \frac{1}{2} (\partial \phi)^2 - \frac{1}{2} \left( \frac{\rho}{M^2} - \mu^2 \right) \phi^2 - \frac{1}{4} \lambda \phi^4 - \frac{\mu^4}{4 \lambda}~.
\ee
Notice the quadratic coupling of the symmetron field $\phi$ to the ambient matter density $\rho$, which leads to a $\mathbb{Z}_2$ symmetry-breaking effective potential,
\be\label{eq:effective_potential}
V_\mathrm{eff}(\phi) = \frac{1}{2} \left( \frac{\rho}{M^2} - \mu^2 \right) \phi^2 + \frac{1}{4} \lambda \phi^4 + \frac{\mu^4}{4 \lambda}~.
\ee
In the absence of matter, $\rho = 0$ and the effective potential has minima at $\phi_\mathrm{min} = \pm v$ with $v \equiv \mu / \sqrt \lambda$. In this case the effective potential reduces to that of a Higgs-like field. Without loss of generality we will consider the positive vacuum value only. The breaking of the $\mathbb{Z}_2$ symmetry can lead to domain walls~\cite{Burrage:2016rkv, Llinares:2018mzl}, which are not expected to be present in the experimental settings that we consider.  When the ambient matter density is above the critical value $\rho_\mathrm{crit} \equiv \mu^2 M^2$, the minimum of the effective potential is located at $\phi = 0$.
If $\rho$ is independent of time then the equation of motion for static field configurations is an elliptic partial differential equation given by
\be
\vec \nabla^2 \phi = \left( \frac{\rho}{M^2} - \mu^2 \right) \phi + \lambda \phi^3~.
\label{eq:EOM}
\ee
The quadratic coupling to matter also leads to a symmetron force experienced by a test particle with mass $m_\mathrm{test}$, given by
\be
\vec a_\mathrm{test} = -\frac{m_\mathrm{test}}{M^2} \phi \vec \nabla \phi~.
\label{eq:test-force}
\ee
 Notice that the force is proportional to the symmetron's local field value---in media with density greater than the critical density, $\rho > \mu^2 M^2$, the ambient field value is driven to $0$ and the symmetron force switches off.
Another remarkable property of the symmetron is that extended objects can be {\it screened}, that is, the symmetron force may not be proportional to the total mass of the object. We review this phenomenon here, for the case of a finite-sized sphere of density $\rho$.  This treatment closely follows that of Ref.~\cite{Hinterbichler:2010es}.

We  compute the symmetron force between a sphere with mass $m_\mathrm{sphere}$ and a test particle with mass $m_\mathrm{test}$.  We first compute the symmetron field profile around the sphere, and then compute the force on the test particle with Eq.~\eqref{eq:test-force}.
The sphere is taken to have radius $R$ and density $\rho(r) = \rho\Theta(R - r)$, where $\rho$ is a constant and $\Theta$ is the Heaviside step function.
We  assume that the sphere is dense enough to restore the $\mathbb{Z}_2$ symmetry, i.e. $\rho > \mu^2 M^2$.  Otherwise, the sphere is not  capable of perturbing the symmetron field away from the vacuum expectation value (VEV) $\phi = v$ and the symmetron force is negligible.

First we solve the equation of motion inside the sphere. Deep inside the sphere we expect $\phi$ to approach the minimum of the effective potential in the symmetric phase. We can therefore approximate the effective potential by a quadratic function located at $\phi = 0$, and find
\be
\phi_\mathrm{in}(r) = A \frac{R}{r} \sinh \left( \mu r \zeta \right)~,
\ee
where we have introduced $\zeta \equiv \sqrt{ \frac{\rho}{\mu^2 M^2} - 1}$. Note that we have set one of the boundary conditions such that $\phi$ does not diverge at the origin. We are left with a normalization factor $A$, which will be determined later.
Next we solve for the field outside the sphere.  We  demand that the field minimizes its potential far away from the sphere: $\phi(r) \to v$ as $r \to \infty$. This time, we linearize the equation of motion about $\phi = v$ and find
\be
\phi_\mathrm{out}(r) = v + B \frac{R}{r} e^{- \sqrt 2 \mu r}~,
\ee
where we have an undetermined constant of integration $B$.
To fix the constants $A$ and $B$, we impose  that the field and its first derivative match at the surface of the sphere $r = R$. We are ultimately interested in the symmetron force exerted on the test particle, so we will only need the constant $B$, for which we find
\be
B = - v e^{\sqrt 2 \mu R} \frac{\mu R\zeta \coth \left(\mu R\zeta  \right) - 1}{\mu R\zeta \coth \left(\mu R\zeta  \right) + \sqrt 2 \mu R}~.
\ee
For the symmetron screening mechanism to work we must assume that the sphere is much denser than the symmetron critical density, i.e. $\rho \gg \mu^2 M^2$. Taking this  limit, the coefficient $B$ becomes
\be
B = - v \left( 1 - \frac{\tanh \sqrt \alpha}{ \sqrt \alpha} \right)~,
\ee
where we have defined the dimensionless parameter $\alpha \equiv \frac{\rho R^2}{M^2}$ for convenience.

Let us now consider a fixed matter coupling $M$, and study the limiting cases of a very small and light sphere versus a very large and dense sphere. In the first limit $\alpha \ll 1$, and we have
\be
B = - v \frac{\alpha}{3}.
\ee
The external field is therefore given by
\be
\phi_\mathrm{out} \approx v - \frac{v \alpha}{3} \frac{R}{r}~.
\ee
The (attractive) symmetron force exerted on a test particle follows from Eq.~\eqref{eq:test-force}, and to leading order in $\alpha$ reads
\begin{equation}
F \approx -\frac{m_\mathrm{test}}{M^2} v  \frac{d }{d r} \left( \phi_\mathrm{out}(r) \right) = - \frac{v^2}{4 \pi M^4} \frac{m_\mathrm{test} m_\mathrm{sphere}}{r^2}~.
\label{eq:force-unscreened}
\end{equation}
Evidently small, light particles couple to each other with a symmetron force proportional to the product of their masses and inversely proportional to the square of the distance between them, exactly as they do in GR.  The relative strength of their interaction, compared to Newtonian gravity, is
\be
\frac{F_\phi}{F_\text{N}} = 2 \left( \frac{v M_\mathrm{Pl}}{M^2} \right)^2~,
\ee
where $M_\mathrm{Pl} \equiv (8 \pi G)^{-1/2}$ is the Planck mass.

Let us now consider the limit of a dense and large sphere, i.e. $\alpha \gg 1$. The integration constant in the external field becomes
\be
B = - v \left( 1 - \frac{1}{\sqrt \alpha} \right)~.
\ee
This time, the symmetron force experienced by the test particle is
\be
F \approx - \frac{v^2}{4 \pi M^4} \frac{m_\mathrm{test} (\lambda_\mathrm{sphere} m_\mathrm{sphere})}{r^2}~,
\label{eq:force-screened}
\ee
where we have defined a {\it screening factor} $\lambda_\mathrm{sphere}$ for the sphere as
\be
\lambda_\mathrm{sphere} \equiv \frac{3 M^2}{\rho R^2}~.
\ee
Notice that the symmetron force is no longer proportional to the total mass of the sphere, but only to a minuscule fraction $\lambda_\mathrm{sphere} \ll 1$ of the sphere's total mass. The force is thus sharply suppressed compared to what one might have guessed based on Eq.~\eqref{eq:force-unscreened} alone. This phenomenon is the essence of screening.

More generally, the symmetron force between two spheres $R_{1, 2}$ involves a screening factor for each sphere.  Combining Eqs.~\eqref{eq:force-unscreened} and \eqref{eq:force-screened}, we have
\be
F = - \frac{v^2}{4 \pi M^4} \frac{(\lambda_1 m_1) (\lambda_2 m_2)}{r^2}~,
\ee
as long as the two spheres can be considered as mutual test objects, i.e. not disturbing the field profile generated by one another,
where the screening factors are
\be
\lambda_i \approx \min \left( \frac{3 M^2}{\rho_i R_i^2}, 1 \right)~,
\label{eq:screening-factor}
\ee
with $i =\{ 1, 2\}$.  The screening factor will play a crucial role in describing the behavior of a small but very dense sphere in the next section.

\subsection{The symmetron force between parallel plates}
\label{sec:parallel_plates}

\begin{figure}[ht!]
\centering
\includegraphics[width = \columnwidth]{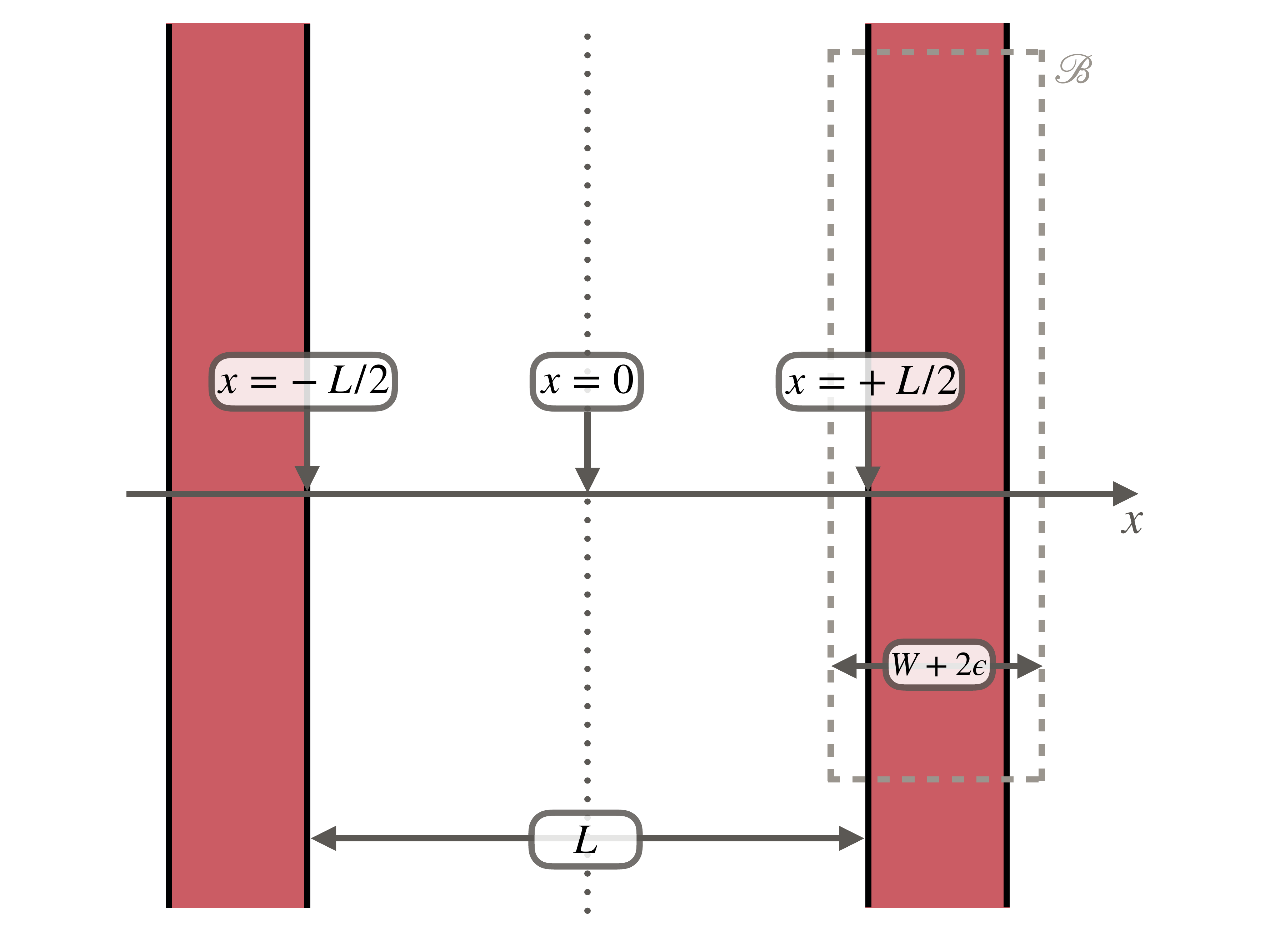}
\caption{\small Parallel-plates configuration. The plates are assumed to have density $\rho \gg \mu^2 M^2$, so that $\phi \approx 0$ everywhere inside.  Everywhere else is assumed to be in perfect vacuum. The force per area on one of the plates is computed by integrating over a boundary that encompasses a section of the plate.}
\label{fig:parallel-plates}
\end{figure}

Let us now discuss another simple density configuration, namely, two infinitely dense parallel plates separated by a distance $L$, and analyze the field profile and the resulting force between the plates; see Fig.~\ref{fig:parallel-plates}. As the density of plates is taken to be infinite, the symmetron field can be taken to be zero inside the plates. Additionally, the exterior solution is unaffected by the separation of the plates, hence we are interested only in the dynamics of the field between the plates.

The parallel-plates configuration is important for two main reasons. First of all, thanks to the planar symmetry of the configuration the symmetron equations of motion can be analytically integrated and the corresponding fifth force can be calculated exactly. This will allow us later to verify the accuracy of our numerical integrator of the symmetron equation of motion, as well as our procedure of numerical force calculation. Besides this technical aspect, as we will see later, the parallel-plates setup can give important insights into the field configuration and the force in the plate-sphere setup.

\subsubsection{Simple approximate treatment}
\label{sec:parallel_plates_approximate}

We start our discussion with an approximate, qualitative treatment of this system. When the separation between the plates is large, $L \gg \mu^{-1}$, the field has sufficient room to roll to the vacuum expectation value. The relevant length scale over which the symmetron field is able to vary significantly is the Compton wavelength in vacuum, $\sim\mu^{-1}$. The gradient energy per unit area of the plate can then be approximated as
\be
\frac{E_\mathrm{roll}}{A} \approx 2\int_0^{\mu^{-1}} \frac{1}{2} (\vec \nabla \phi)^2 \mathrm{d}x \approx \left(\frac{v}{\mu^{-1}}\right)^2 \mu^{-1} = \frac{\mu^3}{\lambda}~,
\ee
where $A$ is the surface area of one of the plates.
Note that due to our convention in Eq.~(\ref{eq:effective_potential}) the true vacuum has a zero potential energy.
As a result the energy of the field configuration for $L \ge \mu^{-1}$ can be approximated as
\be
\frac{E_\mathrm{far}}{A} \approx \frac{\mu^3}{\lambda}~,
\label{energy-plates-far}
\ee
up to an $\mathcal{O}(1)$ constant.
This indicates that the energy stored in the symmetron field does not depend on the plate separation $L$, hence the force between the plates, determined through $F/A = - \frac{\mathrm{d}}{\mathrm{d} L}(E/A)$, is zero: $F_\mathrm{far}/A = 0$.

When the separation $L$ is sufficiently small, $L \lesssim \mu^{-1}$, the field would have to spend a significant amount of gradient energy in order to achieve the true vacuum. As a consequence, the lowest energy configuration is the one in which the field stays in the false vacuum, so $\phi = 0$ everywhere between the two plates. In this case there is no gradient energy, and the total energy per unit plate area is
\be
\frac{E_\mathrm{near}}{A} = \frac{\mu^4}{4 \lambda} L~.
\label{energy-plates-near}
\ee
The force per unit area is, therefore, $F_\mathrm{near}/A = -\mu^4/4 \lambda$.

In summary, according to our qualitative analysis the force between the two plates behaves as
\be
\frac{F}{A} = - \begin{cases}
\frac{\mu^4}{4 \lambda} & \mathrm{if~} L < L_\mathrm{cross}~,\\
0 & \mathrm{if~} L \gtrsim L_\mathrm{cross}~,
\end{cases}
\label{eq:force-planes}
\ee
where $L_\mathrm{cross}$ is a crossover separation below which the field is zero everywhere between the plates. According to the qualitative analysis above we are only able to estimate the order of magnitude of the crossover separation to be $L_\mathrm{cross} \sim \mathcal{O}\left(\mu^{-1}\right)$. In the next subsection we will be able to determine its exact value.

Given these estimates we can now discuss the influence of the quantum corrections to the classical force~\cite{Brax:2018grq}. The quantum interaction between the two plates is not suppressed when the scalar field can be considered as massless in vacuum, i.e. when $\mu L \lesssim 1$. In this case the two plates attract each other with a quantum pressure
\be
\frac{F}{A} \simeq -\frac{\pi^2}{480 L^4}~,
\ee
which is one half of the electromagnetic result as the scalar has only one polarization.
In this regime, the classical  interaction dominates when
\be
\lambda \lesssim \frac{480}{\pi^2} (\mu L)^4~,
\ee
which can be safely taken to be in the range $\lambda \lesssim 1$. In the following, we will always assume that the self-coupling of the symmetron is small enough to
neglect the quantum interaction.

\subsubsection{Improved pressure between plates}
\label{sec:parallel_plates_improved}

The treatment presented above gives a simple estimate of the force.  However, an exact solution is also possible.  As previously mentioned, the symmetries of this density configuration allow the field profile to be found exactly in terms of special functions. Particularly, the solution for the field in the intervening region between the plates is found to be
\be
\phi_\mathrm{int}(x) = v \varphi_0 \cd\left( \mu x \sqrt{1 - \varphi_0^2/2} , \frac{\varphi_0^2}{2 - \varphi_0^2} \right),
\label{eq:parallel_analytic}
\ee
where $\cd(u, m)$ is the Jacobi elliptic function; see, for instance, Refs.~\cite{Brax:2017hna,Burrage:2018xta} for a complete treatment. The parameter $\varphi_0$ ranges between $0$ and $1$ and is fixed by the boundary condition that the field must vanish at the surface of each plate, $\phi_\mathrm{int}(\pm L/2) = 0$. This leads to the condition
\be
0 = \cd\left( \frac{\mu L}{2} \sqrt{1 - \varphi_0^2/2} , \frac{\varphi_0^2}{2 - \varphi_0^2} \right).
\label{eq:phi0_BC}
\ee
Note that $\varphi_0$ is  the value of the field in the middle of the intervening region, normalized by the VEV $v$, i.e. $\varphi_0 = \phi_\mathrm{int}(x = 0)/v$. It is not possible to isolate $\varphi_0$ analytically, although we can note the following:
\begin{itemize}
	\item If $\mu L < \pi$, we must have $\varphi_0 = 0$; see Refs.~\cite{Upadhye:2012rc,Brax:2014zta} for further explanation.
	\item As $\mu L \to \infty$, $\varphi_0 \to 1$.
	\item When $\mu L \gtrsim \pi$, by continuity we can expect $\varphi_0 \ll 1$. We can then Taylor expand Eq.~(\ref{eq:phi0_BC}) and obtain $\varphi_0 \approx \sqrt{\frac{8}{3 \pi}(\mu L - \pi)}$, which we must stop trusting as soon as $\mu L$ is sufficiently larger than $\pi$.
\end{itemize}
Outside the intervening region, the field rises monotonically towards the VEV,
\be\label{eq:field_outside}
\phi_\mathrm{out}(x) = v \tanh \left(\frac{\mu}{\sqrt 2} \left[x - \left(\frac{L}{2} + W\right)\right] \right)~,
\ee
where $W$ is the width of the plate.

We can compute the scalar force on an extended object in the following way~\cite{Hui:2009kc}. The $3$-momentum in a volume of space $\mathcal{V}$ is given by
\be
P_i = \int_\mathcal{V} \mathrm{d}^3 x T_i^{~0}~,
\ee
where $T_{\mu \nu}$ is the total energy-momentum tensor for both the matter and scalar fields, $T_{\mu \nu} = T_{\mu \nu}^\mathrm{matter} + T_{\mu \nu}^\phi$. We choose $\mathcal{V}$ to be large enough so that it encompasses the entire extended object, but small enough that the matter fields dominate the integral. This ensures that the momentum of the volume $\mathcal{V}$ is a good approximation of the momentum of the extended object itself.
The force is the time derivative of the momentum, so we have
\begin{equation}
\dot P_i =  \int_\mathcal{V} \mathrm{d}^3 x \partial_0 T_i^{~0} =  - \int_\mathcal{V} \mathrm{d}^3 x \partial_jT_i^{~j} = - \int_\mathcal{B} \mathrm{d}^2 \sigma_j T_i^{~j}~,
\label{eq:force-general}
\end{equation}
where we have used the conservation of the energy-momentum tensor to go from the first integral to the second, and then changed the volume integral over $\mathcal{V}$ to a surface integral over the outer boundary $\mathcal{B}$ of the volume with an area element $\mathrm{d}^2 \sigma_i$. The matter part of $T_{\mu \nu}$ vanishes on the boundary, so we need to integrate only the energy-momentum tensor of the scalar field. Taking into account the symmetries of the problem, we choose $\mathcal{B}$ to be a rectangular parallelepiped as shown in Fig.~\ref{fig:parallel-plates}, the two sides of which are located at small distances $\epsilon$ from the two surfaces of the plate.
The energy-momentum tensor of the scalar field is
\begin{equation}\label{eq:EMT_tensor}
T^{\phi}_{\mu \nu} = \partial_\mu \phi \partial_\nu \phi + \eta_{\mu \nu} \left( - \frac{1}{2} (\partial \phi)^2 - V(\phi) \right)~,
\end{equation}
and can be computed at any arbitrary point. Obviously, given the symmetries of the configuration, the only non-vanishing component of the force is the $x$-component, which can be written as
\begin{equation}
\dot P_x = - \int_{L/2 - \epsilon}^{L/2 + W + \epsilon} dx \frac{d}{dx} T_{x x}  = T_{xx}(x)\bigg\vert^{L/2 - \epsilon}_{L/2 + W + \epsilon}~
\end{equation}
using the middle integral in Eq.~(\ref{eq:force-general}).

\begin{figure}[ht!]
\centering
\includegraphics[width = \columnwidth]{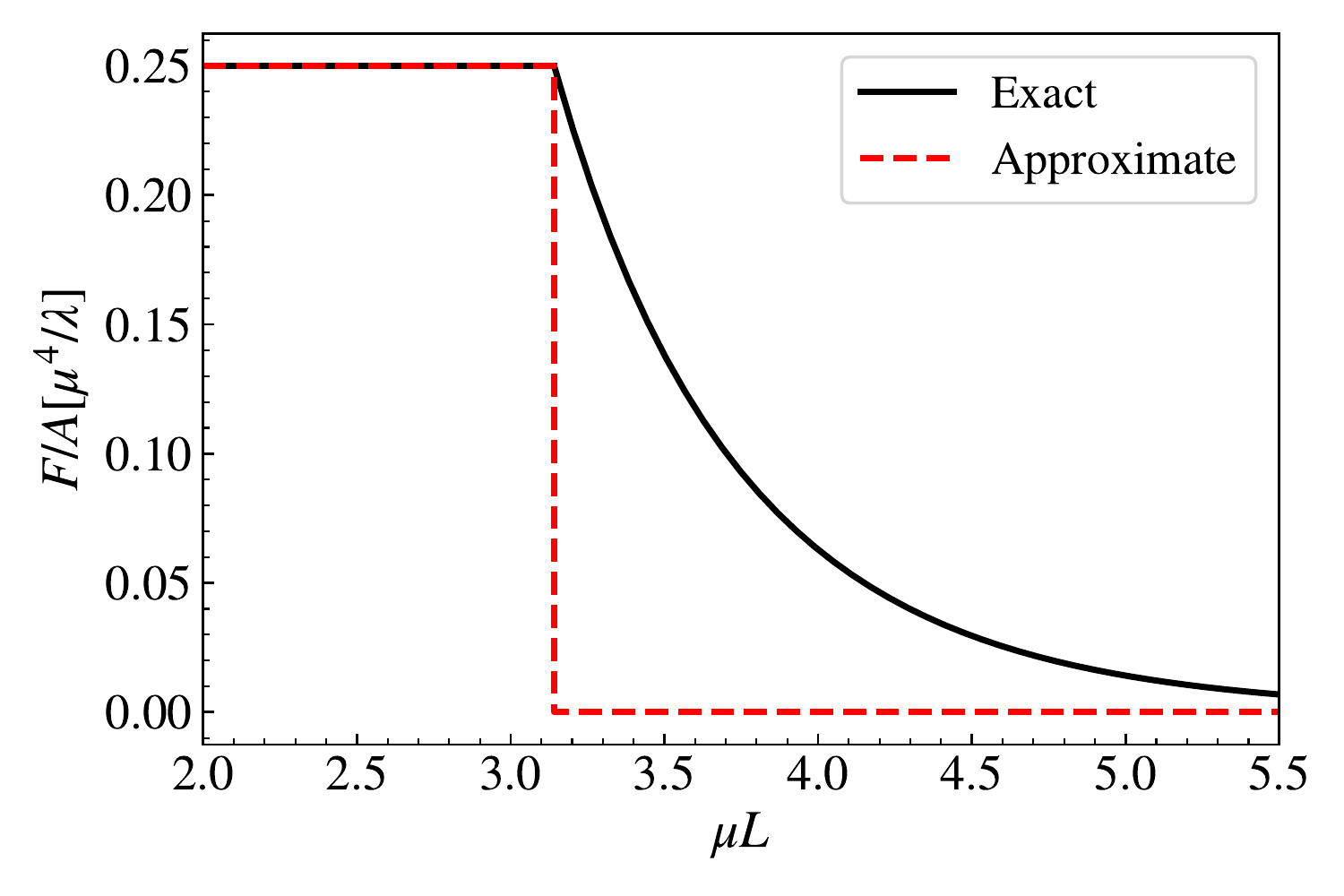}
\caption{\small Pressure exerted by a symmetron fifth force on each plate for an experiment with two parallel plates, computed approximately for small and large separations, as well as through an exact, numerical method. The pressure is constant for plate separations $L < \pi \mu^{-1}$ and then decreases exponentially for larger separations.}
\label{fig:plates_pressure}
\end{figure}

Since the scalar field vanishes on the surface of the plate, we obtain
\begin{equation}
\dot P_x = \frac{1}{2} \left[\phi_\mathrm{int}'\left(\frac{L}{2}\right)\right]^2 -  \frac{1}{2} \left[\phi_\mathrm{out}'\left(\frac{L}{2} + W\right)\right]^2~,
\end{equation}
where a prime denotes a derivative with respect to $x$. Note that the right-hand side of this equation is equal to the difference between the boundary values of the stress-energy tensor, as expected from the last expression in Eq.~(\ref{eq:force-general}).
We now use the solutions of the field given by Eqs.~(\ref{eq:parallel_analytic}) and (\ref{eq:field_outside}), as well as the identities
\begin{align} \nonumber
\frac{d}{dx} \tanh(x) &= 1 - \tanh(x)^2~, \\
\left( \frac{d }{dx} \cd(x, m) \right)^2 &= (1 - \cd(x, m)^2)(1 - m^2 \cd(x, m)^2)~,
\end{align}
to calculate the force between the plates. We obtain a force per area $A$
\be
\frac{F}{A} = \frac{\mu^4}{4 \lambda} \left( \varphi_0^2 (2 - \varphi_0^2) - 1 \right)~.
\label{eq:exact-pressure}
\ee
When the distance between the two plates is small, $\mu L < \pi$, then $\varphi_0 = 0$ and $F/A = -\frac{\mu^4}{4 \lambda}$, as we obtained in our qualitative estimation of the force in the previous subsection. When the distance is large, $L \to \infty$, then $\varphi_0 \to 1$ and $F/A \to 0$. For general values of $L$, we solve Eq.~(\ref{eq:phi0_BC}) numerically, using Newton's method. We plot in Fig.~\ref{fig:plates_pressure} the exact pressure $F/A$ as a function of $\mu L$, as well as the qualitative estimates for small and large distances. Note that the estimated values for small distances, $\mu L < \pi$, are in perfect agreement with the exact values, while they are significantly smaller for intermediate distances, $\pi<\mu L \lesssim 5.5$; the estimated and exact values agree for $L \to \infty$. This is an important observation with implications for calculating the constraints one may want to impose on the symmetron fifth forces using parallel-plates experiments.

\section{Plate and sphere I: analytic results}
\label{sec:analytic}

In this section, we turn to the problem of computing the classical scalar force between an infinite plate and a finite-radius sphere. The reason for considering this configuration is that Casimir experiments are often built using  a plate-sphere system~\cite{Lamoreaux:1996wh, Decca:2005yk, Chen:2014oda}, as this setup eliminates experimental uncertainty related to proper alignment of two plates.

\begin{figure}[ht!]
\centering
\includegraphics[width = \columnwidth]{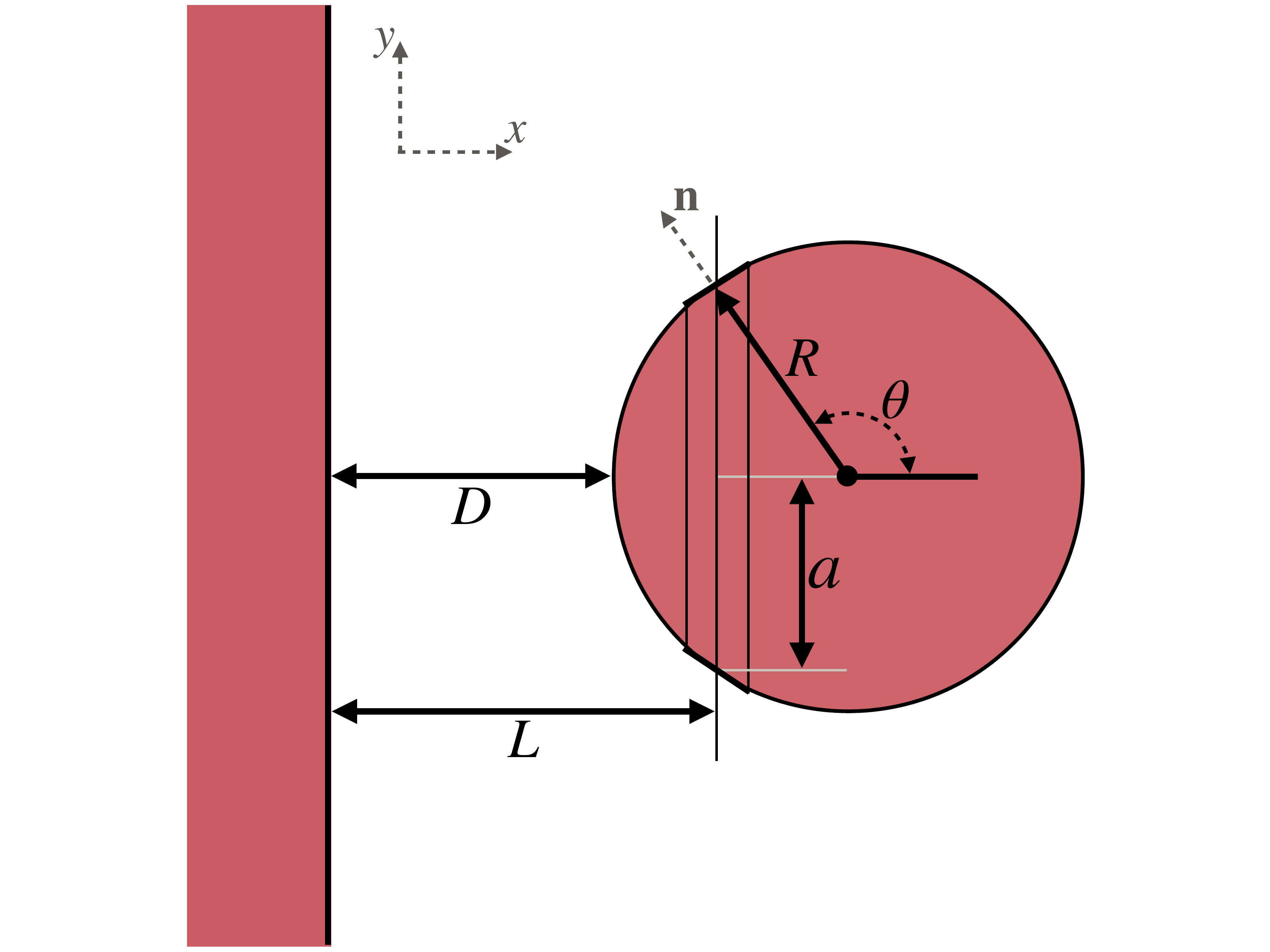}
\caption{\small Plate-sphere setup.  The plate is assumed to be infinitely large and separated by a distance $D$ from the surface of a sphere of radius $R$. We have also illustrated various geometric quantities that will be useful. The plate and the sphere are located in vacuum and both are assumed to be sufficiently dense so that $\phi \approx 0$ everywhere inside the objects.}
\label{fig:plate-sphere}
\end{figure}

We assume that the surface of the infinite plate is located at $x = 0$ with a sphere of radius $R$ facing it, and that the nearest distance between the plate and the sphere is $D$; see Fig.~\ref{fig:plate-sphere}. Although a given experimental setup is likely to have a fixed radius $R$ and to probe only a few different separations, our aim in this paper is to compute the symmetron force for any combination of $R$ and $D$.  This, along with the scale invariance of our solutions, will allow our results to be applicable to any system that resembles a plate-sphere configuration, ranging from microscopic to astrophysical scales. Furthermore, our general approach will enable future work to identify optimal configurations for new Casimir experiments, in order to probe the most interesting regions of the parameter space.

Due to the nonlinearity of the symmetron's equation of motion, it is impossible to find exact analytic solutions for the field in this configuration. We must therefore work with numerical solutions or analytic approximations in different regimes. In this section, we present approximate solutions for the scalar force in two regimes: (i) large spheres with $R \gg \mu^{-1}$, and (ii) small spheres with $R \ll \mu^{-1}$.  In the next section, we use numerical solutions to verify these approximations, as well as to provide solutions for $R$ of $\mathcal{O}(\mu^{-1})$.

\subsection{Large sphere}

We first analyze the plate-sphere system in the limit that the sphere radius is much larger than the symmetron's Compton wavelength, i.e. $R \gg \mu^{-1}$. Our approach is similar to the \textit{proximity approximation} that is often employed in calculations of the quantum Casimir force between objects with complex geometries; see, for example, Refs.~\cite{Krause:2007zz, Brax:2007vm}.

We consider infinitesimally thin ring-like elements of the sphere parallel to the plate, and use the analytic formula for the parallel-plate pressure calculated in the previous section to model the pressure exerted on the element by the plate. Let us consider a ring placed at a distance $L = D + R + R \cos \theta$ from the surface of the plate, where $\theta$ is the standard polar angle defined around the $x$-axis (see Fig.~\ref{fig:plate-sphere}). The contribution of this ring to the total force experienced by the sphere is approximated as
\be
\mathrm{d}F = P(L) (-\cos \theta) \mathrm{d}A  ~,
\ee
where $P(L)$ is the pressure between two parallel plates separated by the distance $L$, and $dA$ is the area of the infinitesimal ring element. The factor $- \cos \theta$ takes the $x$-component of the force on the ring or corresponds to the cross-sectional area of the sphere directly perpendicular to the plate. The area of the ring is $\mathrm{d}A = 2\pi a (R d \theta)$, where $a = R \sin \theta$ is the vertical distance from the infinitesimal ring to the line connecting the plate and center of the sphere.

The total symmetron force exerted on the sphere is then obtained by integrating over the hemisphere nearest the plate and is given by
\begin{equation}
F = - 2 \pi R^2 \int_{\pi / 2}^\pi \mathrm{d} \theta ~ \sin \theta \cos \theta P(D + R + R \cos \theta)~.
\end{equation}
What remains is to specify the pressure between the parallel plates $P(L)$. In Sec.~\ref{sec:parallel_plates}, we found two different expressions for that. The first one is given in Eq.~\eqref{eq:force-planes}, which is approximate but simple. We will refer to this as $P_\mathrm{approx}(L)$.  We also found the exact pressure between two plates, given by Eq.~\eqref{eq:exact-pressure}, which we will refer to as $P_\mathrm{exact}(L)$. Although the latter has the advantage of being exact, it does not admit a closed-form solution.
Employing the former for now, we note that the pressure is a step function, so only a spherical cap within a distance $\pi / \mu$ from the plate contributes to the pressure, and $P_\mathrm{approx}(L)$ is constant over this area.
We therefore compute the attractive force on the spherical cap, which has a height $h = \pi / \mu - D$ and radius $a^2 = R^2 - (D + R - \pi / \mu)^2$. The cap has cross-sectional area $\pi a^2 = \pi (\pi / \mu - D)(2 R - (\pi / \mu - D))$, so the force is
\be
F = -\frac{ \mu^4}{4 \lambda} \pi (\frac{\pi}{\mu} - D)(2 R - (\frac{\pi}{\mu} - D)) \quad \mathrm{if} \quad D < \frac{\pi}{\mu} < D + R~.
\ee
When no part of the sphere is within $\pi / \mu$ of the plate, the force is zero:
\be
F = 0 \quad \mathrm{if} \quad D  > \frac{\pi}{\mu}~.
\ee
When all of the near hemisphere is within $\pi / \mu$ of the plate, we have
\be
F = -\frac{\pi R^2 \mu^4}{4 \lambda} \quad \mathrm{if} \quad D + R < \frac{\pi}{\mu}~.
\ee
In summary, the force is
\be
F = \begin{cases}
-\frac{\mu^4}{4 \lambda} \pi R^2 & D < \frac{\pi}{\mu} - R~, \\
-\frac{\mu^4}{4 \lambda}\pi \left(\frac{\pi}{\mu} - D\right)\left(2R + D - \frac{\pi}{\mu}\right) & \frac{\pi}{\mu} - R < D < \frac{\pi}{\mu}~,\\
0 & D > \frac{\pi}{\mu}~.
\end{cases}
\label{eq:plate-sphere-force}
\ee
We remind the reader that we are working in the regime of large spheres, $R \gg \mu^{-1}$, so the first case is included only for completeness.

To obtain a more exact result, we could instead have used $P_\mathrm{exact}(L)$, given by Eq.~\eqref{eq:exact-pressure}.  In this case, it is not possible to simplify the expression, and we must instead rely on numerical integration.  That is, we integrate
\begin{align}
F &= - 2 \pi R^2 \int_{\pi/2}^\pi d\theta ~ \sin \theta \cos \theta ~ P(D + R + R\cos\theta)~,
\label{improved-force-plate-sphere}
\end{align}
where the pressure $P$ is given by Eq.~\eqref{eq:exact-pressure}.  A comparison between these two expressions for the symmetron force on a large sphere $R = 10 \mu^{-1}$ is given in Fig.~\ref{fig:large_and_small_R}.

\subsection{Small sphere}

Let us now calculate the symmetron force between a plate and a small sphere, i.e. $R \ll \mu^{-1}$. In this limit, we may safely neglect the backreaction of the sphere on the symmetron field profile sourced by the plate alone.\footnote{Note that the backreaction, neglected here, may lead to large corrections at very small distances, as explored in Ref.~\cite{Ogden:2017xeo}.} This means that the force on  the sphere will be proportional to that of a test particle. The constant of proportionality is the screening factor $\lambda_\mathrm{sphere}$, which is given by Eq.~\eqref{eq:screening-factor}. The force on the sphere is therefore
\be
\vec F = - \frac{ \lambda_\mathrm{sphere} m_\mathrm{sphere} }{M^2} \phi \vec \nabla \phi~,
\label{eq:force-screened2}
\ee
where $m_\mathrm{sphere}$ is the mass of the sphere, and $\phi$ is the field profile generated by the plate alone. Recall that the latter is simply given by
\be\label{eq:plane_field_profile}
\phi(x) = v \tanh \left( \frac {\mu x}{\sqrt{2}}\right)~.
\ee
Before putting all these ingredients together let us also recall that we are working in the limit $\rho R^2 \gg M^2$, and as such Eq.~\eqref{eq:screening-factor} indicates that $\lambda_\mathrm{sphere} \to 0$. However, in this limit the mass of the sphere diverges, and as a result the prefactor in Eq.~(\ref{eq:force-screened2}) is a non-negligible quantity given by
\be
\frac{\lambda_\mathrm{sphere} m_\mathrm{sphere}}{M^2} = 4 \pi R.
\label{screening-factor}
\ee
Putting all these together, Eqs.~(\ref{eq:plane_field_profile}) and (\ref{eq:force-screened2}) give the attractive force
\be
F = - \frac{4 \pi v^2 \mu R}{\sqrt 2}  \tanh \left( \frac {\mu x}{\sqrt{2}}\right) \mathrm{sech}^2 \left( \frac {\mu x}{\sqrt{2}}\right)~,
\label{force-small-sphere}
\ee
where $x$ is the distance from the surface of the plate to the center of the sphere.
\begin{figure*}
\centering
\includegraphics[width = \columnwidth]{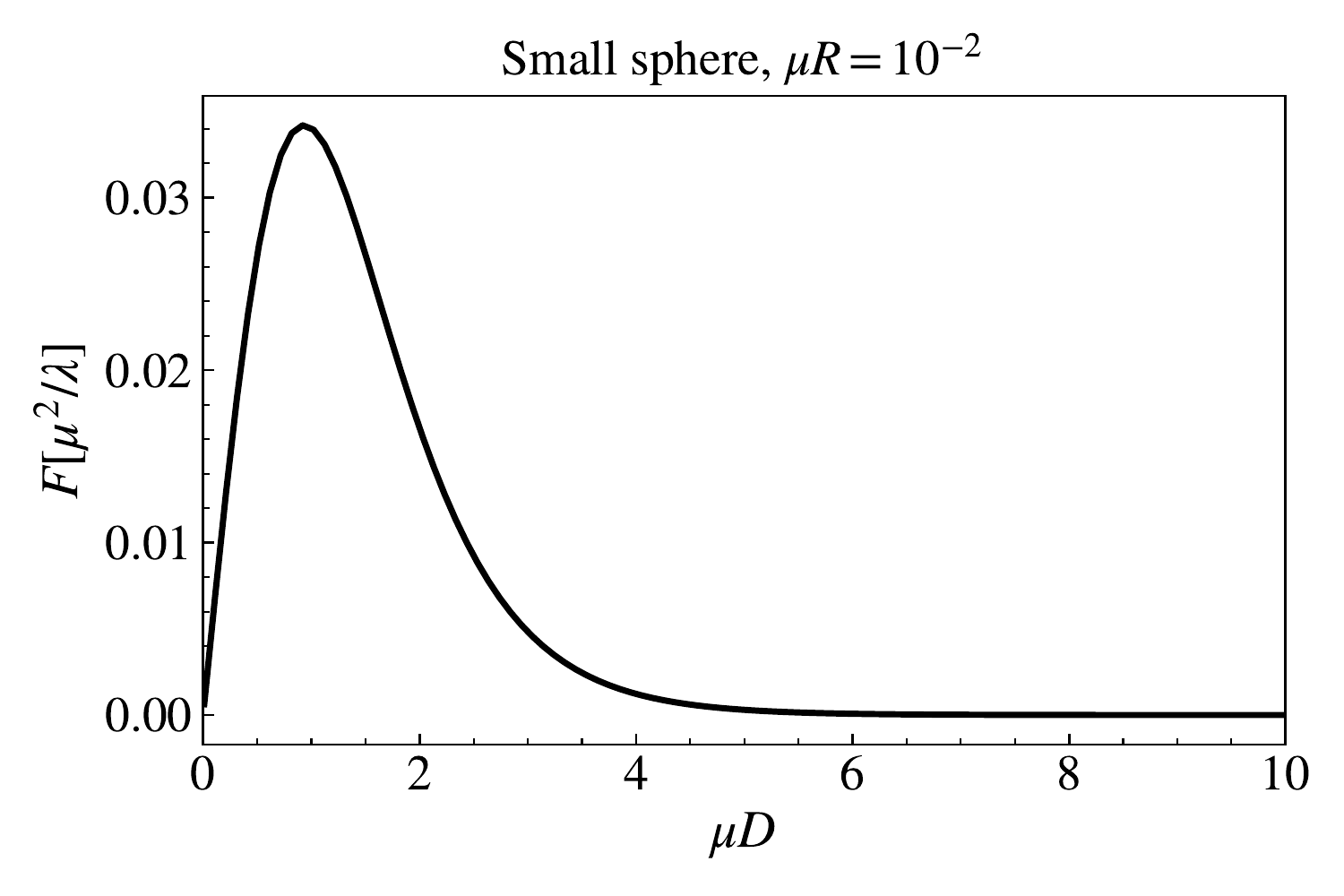}
\includegraphics[width = \columnwidth]{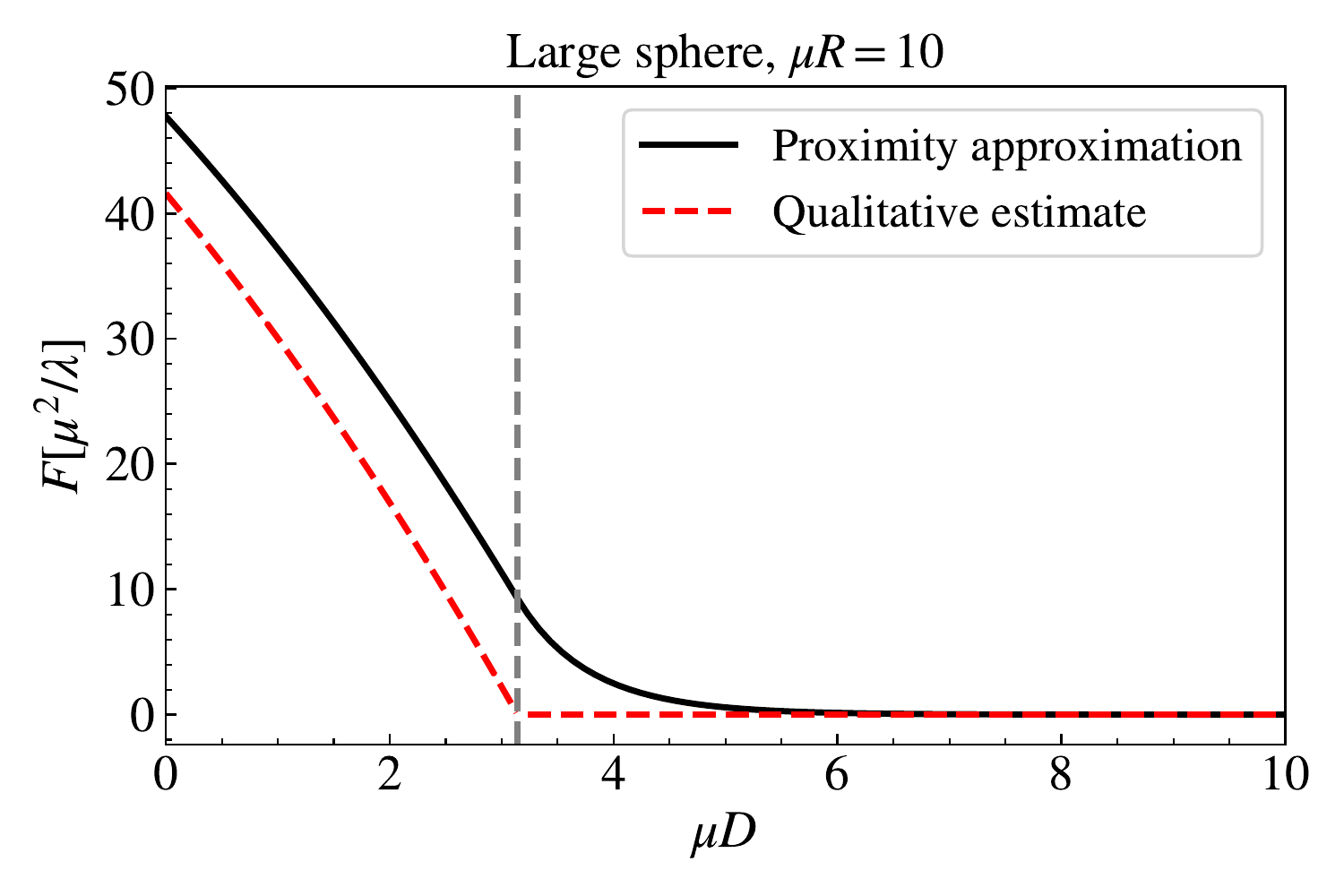}
\caption{\small Comparison of approximate symmetron forces for small and large spheres in a plate-sphere setup. {\bf Left panel:} The force is computed for small spheres with $\mu R \ll 1$ using Eq.~\eqref{force-small-sphere}. Here, we have considered a screening factor for the sphere and neglected the backreaction of the sphere on the symmetron field profile generated by the plate.  {\bf Right panel:} The force is computed for large spheres with $\mu R \gg 1$ using the proximity force approximation, given by Eq.~\eqref{improved-force-plate-sphere}. A qualitative estimate of the force is also plotted for comparison, given by Eq.~\eqref{eq:plate-sphere-force}.  The vertical line indicates the distance $\mu D=\pi$ beyond which the qualitative estimate vanishes.}
\label{fig:large_and_small_R}
\end{figure*}

The approximate forces we obtained for large and small spheres are plotted in Fig.~\ref{fig:large_and_small_R} as functions of the distance between the plate and the sphere. Notice that in the case of the small sphere the force vanishes as the sphere approaches the plate. Intuitively, this may be understood as a consequence of the symmetron's quadratic coupling to matter: near the plate the ambient field value is driven towards zero, so the coupling between the symmetron and matter becomes weaker and the force is diminished.

\section{Plate and sphere II: numerical results}
\label{sec:numerical}

Beyond highly symmetric configurations it is impossible to find exact solutions of Eq.~\eqref{eq:EOM}.  In the previous section, we detailed two useful limiting cases: very large spheres with $R \gg \mu^{-1}$, and very small spheres with $R \ll \mu^{-1}$.  In these cases it was possible to exploit the hierarchy of scales to make analytic progress.

In this section, we turn to the intermediate regime, where $R \approx \mu^{-1}$.  We are unable to exploit any such hierarchy of scales in this case, so we must resort to numerical integration of the symmetron's equation of motion. Our results allow us to bridge the gap between the two limiting cases we explored in the previous section. We use a commercial solver,\footnote{We used Matlab's Partial Differential Equation Toolbox, version R2019b, similar to what has been done in Refs.~\cite{Chiow:2018lze, Chiow:2019fti}.} which relies on the {\it finite-element} method. For some of the cases, we have checked the solutions using our own numerical code based on the {\it Newton-Gauss-Seidel relaxation algorithm}, which is a standard method widely used in the studies of screened modified gravity (see Ref.~\cite{Elder:2016yxm} for a description of the algorithm, as well as Refs.~\cite{Jaffe:2016fsh,Sabulsky:2018jma}). This has allowed us to assess the validity of the solutions obtained by two different solvers.

As before, we will assume that the sphere and the plate are infinitely dense, and that the matter density elsewhere is zero. The former assumption allows us to set $\phi = 0$ everywhere inside the objects, which gives us the boundary condition $\phi = 0$ at the surfaces of the plate and the sphere. We additionally take the component of $\nabla \phi$ normal to the simulation box edges to vanish. In all cases, we have checked that our simulation box size, $L_\mathrm{box}$, is large enough, so any boundary effects do not alter the field profile around the sphere.

At first it may appear that we have four parameters that specify a given configuration, $\mu$, $\lambda$, $R$ and $D$. However, the static symmetron's equation of motion is invariant under simultaneous rescalings of the field and coordinates. This rescaling symmetry makes it possible to eliminate the $\mu$ and $\lambda$ parameters from the equation of motion, significantly reducing our computational task.
To see this explicitly, we define a new (dimensionless) field variable $\varphi \equiv \phi / v$, in terms of which the equation of motion in the absence of matter ($\rho = 0$) is given by
\be
\nabla^2 \varphi = - \mu^2 \varphi + \mu^2 \varphi^3~.
\ee
Next, let us rescale all spatial dimensions by $\mu$ so that our new dimensionless distance coordinate is $\hat x \equiv \mu x$. The equation of motion now becomes
\be
\hat \nabla^2 \varphi = - \varphi + \varphi^3~,
\label{dimensionless-eom}
\ee
where the hat indicates a derivative with respect to the new dimensionless coordinate $\hat x$.  We are left with just two parameters that completely define the plate-sphere system. These parameters are the rescaled radius of the sphere, $\hat R = \mu R$, and the rescaled distance between the surface of the sphere and the plate, $\hat D = \mu D$. We will solve the equation numerically in this form and only later will we reintroduce the parameters $\mu$ and $\lambda$.

Once we have the field profile, we can compute the scalar force on the sphere using Eq.~\eqref{eq:force-general}. In particular, we perform a surface integral of the energy-momentum tensor, over a spherical surface taken to be slightly larger than the sphere itself, i.e. with a radius $\hat{r} = \hat{R} + \epsilon$. Since we will ultimately use interpolated values of the field, henceforth we drop the $\epsilon$ and take the radius of the integral to be $\hat{r} = \hat{R}$. Then the analogue of Eq.~\eqref{eq:force-general} for the rescaled $\{\hat{x}, \hat{y}\}$ coordinates gives
\begin{align} \nonumber
\hat{F} & \equiv - \int_B \mathrm{d}\hat{A} ~\hat{n}_j \hat{T}_i^{~j}\\
&=  (- 2 \pi \hat R^2)\!\int_0^\pi\!\sin \theta d\theta\!\left( \cos \theta \hat T_{x x}(\hat R, \theta) + \sin \theta \hat T_{x y} (\hat R, \theta) \right),
\label{numerical-force}
\end{align}
where $\hat{T}_{\mu \nu}$ is obtained by rewriting Eq.~(\ref{eq:EMT_tensor}) in our rescaled variables.
Note also that $\hat{\textbf{n}} = \cos \theta \hat{\textbf{x}} + \sin \theta \hat{\textbf{y}}$ is the normal vector to the sphere, see Fig.~\ref{fig:plate-sphere}.
Given a numerical solution $\varphi(\hat{x}, \hat{y})$, we can interpolate the values of $\hat{T}_{\mu \nu}$ in the integrand, and obtain the dimensionless scalar force on the sphere. Notice that the force is perpendicular to the plate as contributions to the force from opposite sides of the sphere, which are directed parallel to the plate, cancel each other out.
It is then straightforward to reintroduce the factors of $\mu$ and $\lambda$ back and obtain the physical scalar force $F$,
\begin{equation}\label{eq:physical_force}
F = \frac{\mu^2}{\lambda}\hat{F}.
\end{equation}

Our aim is to  solve Eq.~\eqref{dimensionless-eom} numerically  throughout  the region $0.1 \lesssim \hat R \lesssim 10$, as larger or smaller values are covered by the approximate methods detailed in the previous sections. We shall see that our numerical results are well-described by those approximations in the appropriate limits. Let us also mention that our primary interest is in relatively small plate-sphere separations, typically $\hat D \lesssim 10$, as the force is exponentially suppressed with distance, so larger separations are not relevant for fifth force tests.

We set the edges of the simulation box to be at least $\hat{H}$ Compton wavelengths away from the surface of the sphere, so the dimensions of the simulation box are $(\hat{D} + 2\hat{R} + \hat{H}) \times (\hat{R} + \hat{H})$---note that our simulations are performed in cylindrical polar coordinates, hence in the perpendicular direction the size of the simulation box is $(\hat{R} + \hat{H})$ and not $2(\hat{R} + \hat{H})$. We have chosen $\hat{H} = 5$ as a balance between being a small enough box size to have good enough resolution, and it being large enough to be able to safely neglect the effects of the simulation box boundaries on the field profile near the sphere. The size of the mesh is set sufficiently fine that further increases do not change the solution by more than $1\%$.
\begin{figure}[hpt!]
\centering
\includegraphics[width = \columnwidth]{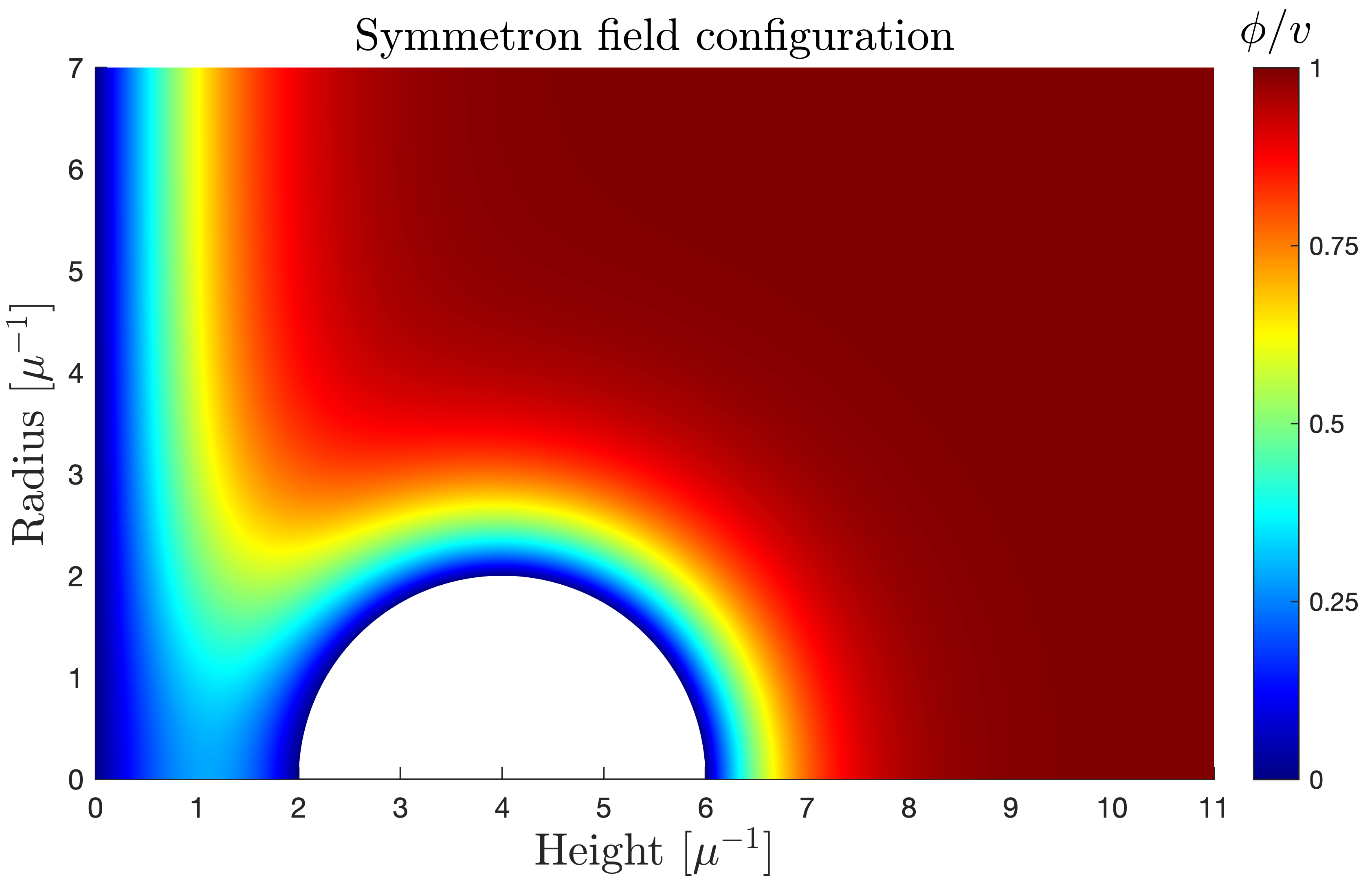}
\caption{\small Typical numerical solution of the symmetron field around the plate and sphere setup for $D = R = 2 \mu^{-1}$. The color indicates the field value as a fraction of the VEV.  The simulation is performed in cylindrical coordinates, in order to exploit the axisymmetry of the problem.  From the field and its gradient, the force may be computed via Eq.~\eqref{numerical-force}.}
\label{fig:plate-sphere-field}
\end{figure}

\begin{figure*}[ht!]
\centering
\includegraphics[width = 0.75\textwidth]{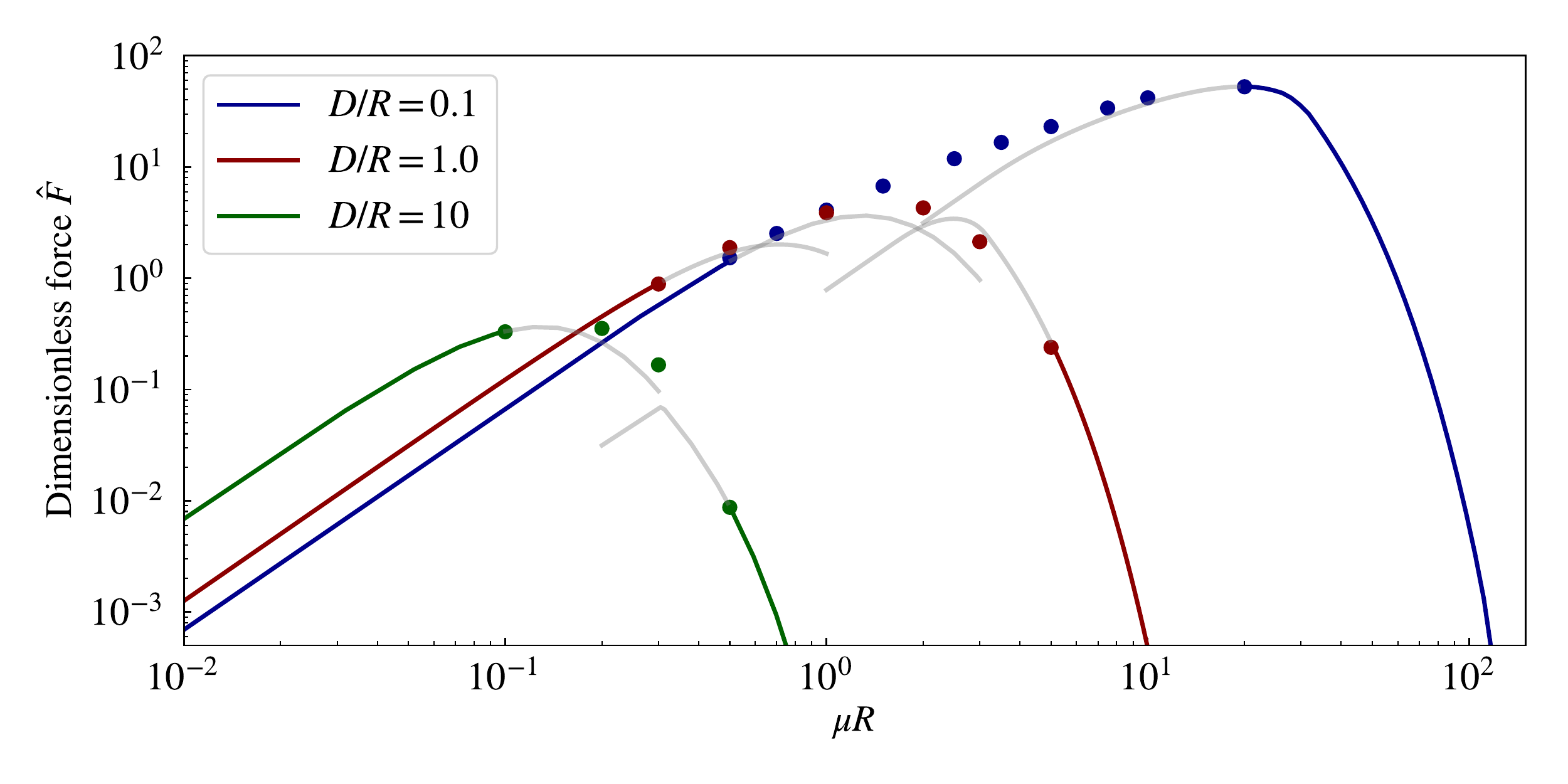}
\caption{\small Symmetron force $\hat F$ between a sphere of radius $R$ and an infinite plate a distance $D$ away. The force is given as a dimensionless number $\hat F = \lambda F / \mu^{2}$, and is a function of the rescaled radius of the sphere $\hat R=\mu R$ and the value of $D/R$. The solid curves show the analytic approximations corresponding to small spheres (left-most curves) and large spheres (right-most curves). The dots are results of numerical integration of the full, nonlinear equation of motion of the field, with the force being computed using the exact (numerically calculated) field profile. The pale gray curves represent the analytic results in the regimes where they are no longer valid, hence should be replaced with numerical results. The \textit{leftmost} and \textit{rightmost} points of each color are in perfect agreement with the corresponding analytic approximations. These points, therefore, mark the upper and lower validity regimes of the analytic approximations.}
\label{fig:force-plate-sphere}
\end{figure*}

For a range of configurations with $D/R \in [0.1, 10]$ we have numerically solved Eq.~\eqref{dimensionless-eom} for different values of $\hat R$. The smallest and largest values of $\hat R$ are determined by identifying the validity bounds of the small-sphere and large-sphere approximations. For example, for $D/R = 0.1$, numerical simulations are perfomed in the range $\hat R \in [0.5, 20]$. The resulting force has been computed with Eq.~\eqref{numerical-force}. In this range one can neglect the quantum interactions between the sphere and the plate as long as the self-coupling $\lambda $ is small enough, typically less than a coupling of order one. Fig.~\ref{fig:plate-sphere-field} depicts an example of the field profile as a result of our numerical integration.  Our results for the force are plotted in Fig.~\ref{fig:force-plate-sphere}. From Fig.~\ref{fig:force-plate-sphere}, we clearly see that the analytic approximations developed in Sec.~\ref{sec:analytic} match the numerical results well for $\mu R \lesssim 0.1$ (small spheres) and $\mu R \gtrsim 10$ (large spheres).

\section{Forecast constraints on symmetron's parameters}
\label{sec:forecast}

In this section, we turn to the task of using Casimir force sensors to place bounds on the symmetron's free parameters. Currently, there exists a precise Casimir force sensor, employing a plate-sphere configuration, which could be used to constrain the symmetron's mass and couplings. The sensor consists of a sphere of radius $R = 150$ $\mu$m, held at a distance $D_\mathrm{near} = 15$ $\mu$m from a circular plate. The plate has a pattern of $50 ~\mu$m deep rectangular trenches of length and width $200$ $\mu$m cut into it.  The plate is rotated so that the trenches pass under the sphere.  The entire setup is inside a vacuum chamber of height $40$ cm and diameter $45$ cm.
As the plate rotates, the horizontal distance $D$ between the sphere and the plate alternates between $D_\mathrm{near} = 15$ $\mu$m and $D_\mathrm{far} = 65$ $\mu$m.  The experiment is very similar to, but more sophisticated than that of Ref.~\cite{Chen:2014oda}, the details of which will appear shortly.
Provided all known interactions have been ruled out, when the measured force is null within the experimental error (currently 0.2 $f$N at the 95\% confidence level~\cite{Chen:2014oda}),
\be
\delta F = F_\mathrm{near} - F_\mathrm{far}  = 0 \pm 0.2 {~\rm fN}
\label{eq:expt_bound}
\ee
yields the fifth force constraints.

Among known interactions, the difference in the Newtonian gravitational force between the near and far configurations is too small to be detected ($\sim 1$ zN), the electrostatic interaction can be controlled to better than 10 aN, and the Casimir interaction itself has been measured to be much smaller than 0.2~fN for separations $D > 12 ~\mu$m.

Ideally, one would then use the results from Secs.~\ref{sec:analytic} and \ref{sec:numerical} to calculate the difference in the symmetron force between these two configurations $\delta F_\phi$, allowing one to place constraints on the symmetron's parameters.
However, there is a complication due to the finite size of the trenches.  In order to place accurate constraints, one would either need the trench walls to be so far away from the sphere that they are negligible, or one would need to perform detailed 3D simulations of the setup.
The focus of this work is on the plate-sphere system, and accurately describing the effects of the trenches is beyond the scope of the present paper. Therefore, we shall use the results of the pure plate-sphere system, derived in Secs.~\ref{sec:analytic} and \ref{sec:numerical}, to produce {\it forecasts} for an as-yet unperformed experiment that has an identical setup, except with trenches that are much wider and longer, so that the trench walls could be safely neglected.  The system could then be treated as a pure plate and sphere, oscillating between two separations $D_{\rm near}$ and $D_{\rm far}$.

Depending on the size of the parameter $\mu$, we shall employ different approximations for the symmetron force between the plate and the sphere.  As we saw in Sec.~\ref{sec:numerical}, the proximity approximation given by Eq.~(\ref{improved-force-plate-sphere}) works well for $\mu \gtrsim 10 R^{-1}$.  Likewise, we saw that the small screened sphere approximation, given by Eq.~(\ref{force-small-sphere}), may be trusted when $\mu \lesssim 0.1 R^{-1}$.  In between, we must rely on numerical solutions, which interpolate between these two regimes, as described in the previous section.

Before presenting the forecast constraints let us note that we are able to place constraints only in the regime where $\mu \gtrsim R_\mathrm{vac}^{-1}$, with $R_\mathrm{vac}$ being the smallest interior dimension of the vacuum chamber.  If this condition is not met then the symmetron field stays in the false vacuum $\phi = 0$ everywhere inside the vacuum chamber, just as we saw in Sec.~\ref{sec:parallel_plates}. Consequently, there is no fifth force and it is impossible to place constraints. Additionally, very large values of $\mu$ correspond to very large values of $\hat{R}$ and $\hat{D}$ (with $\hat{D}/\hat{R}$ being fixed), which, according to Fig.~\ref{fig:force-plate-sphere}, correspond to vanishingly small $\hat{F}$. Intuitively, in this regime the range of the force is much shorter than the distance between the plate and the sphere, and the force is exponentially suppressed. As a result, very large values of $\mu$ are also not well constrained. Moreover, as we mentioned above, the experiments are able to report only differential force measurements, i.e. force differences between two sphere placements. As we see in Fig.~\ref{fig:force-plate-sphere}, there is an intermediate value of $\mu$ for which any of the two curves (corresponding to different values of $\hat{D}/\hat{R}$) intersect each other. The differential signal in the vicinity of this value of $\mu$ is therefore tiny, and hence these particular values of $\mu$ cannot be constrained.
\begin{figure*}
\centering
\includegraphics[width=\columnwidth]{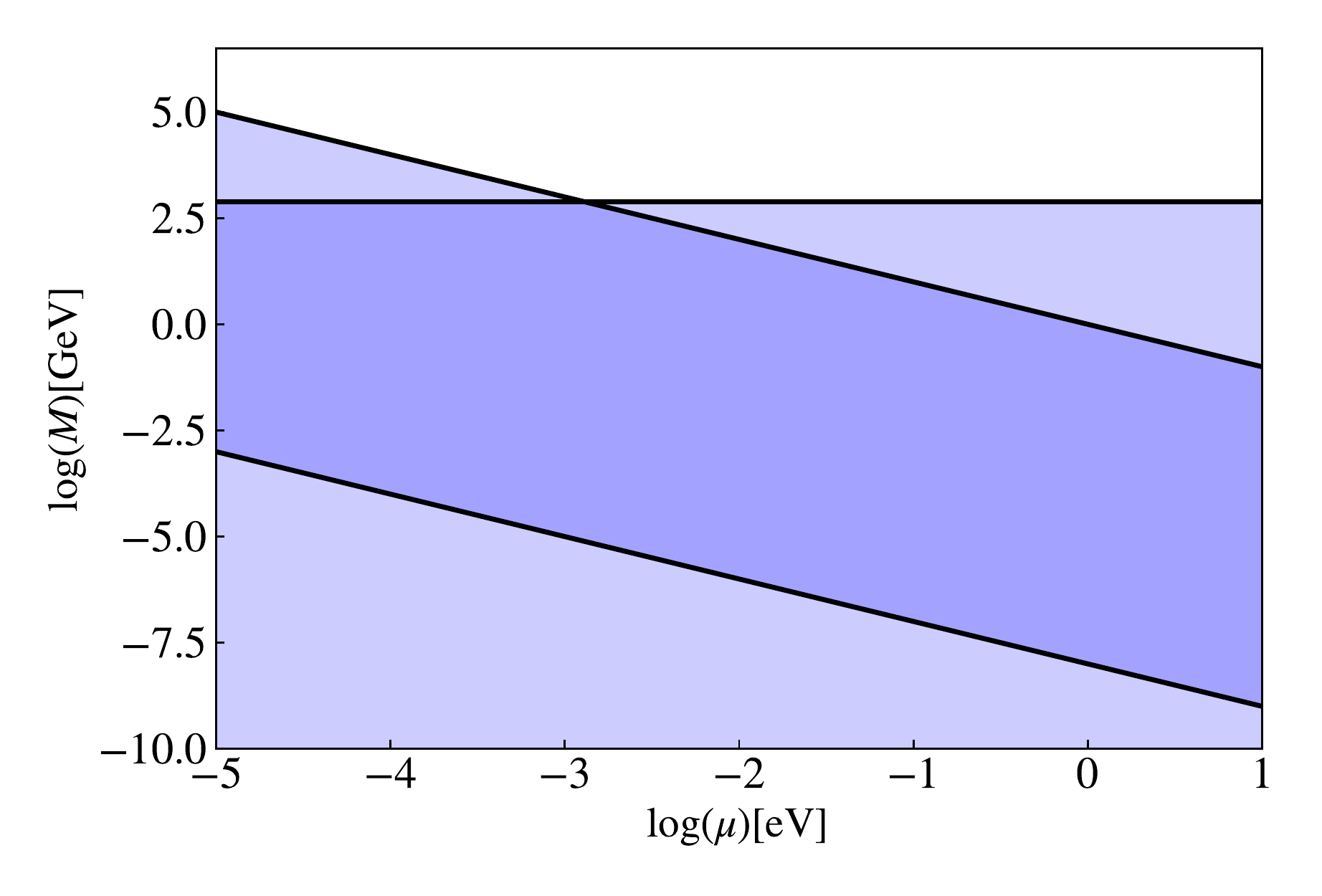}
\includegraphics[width=\columnwidth]{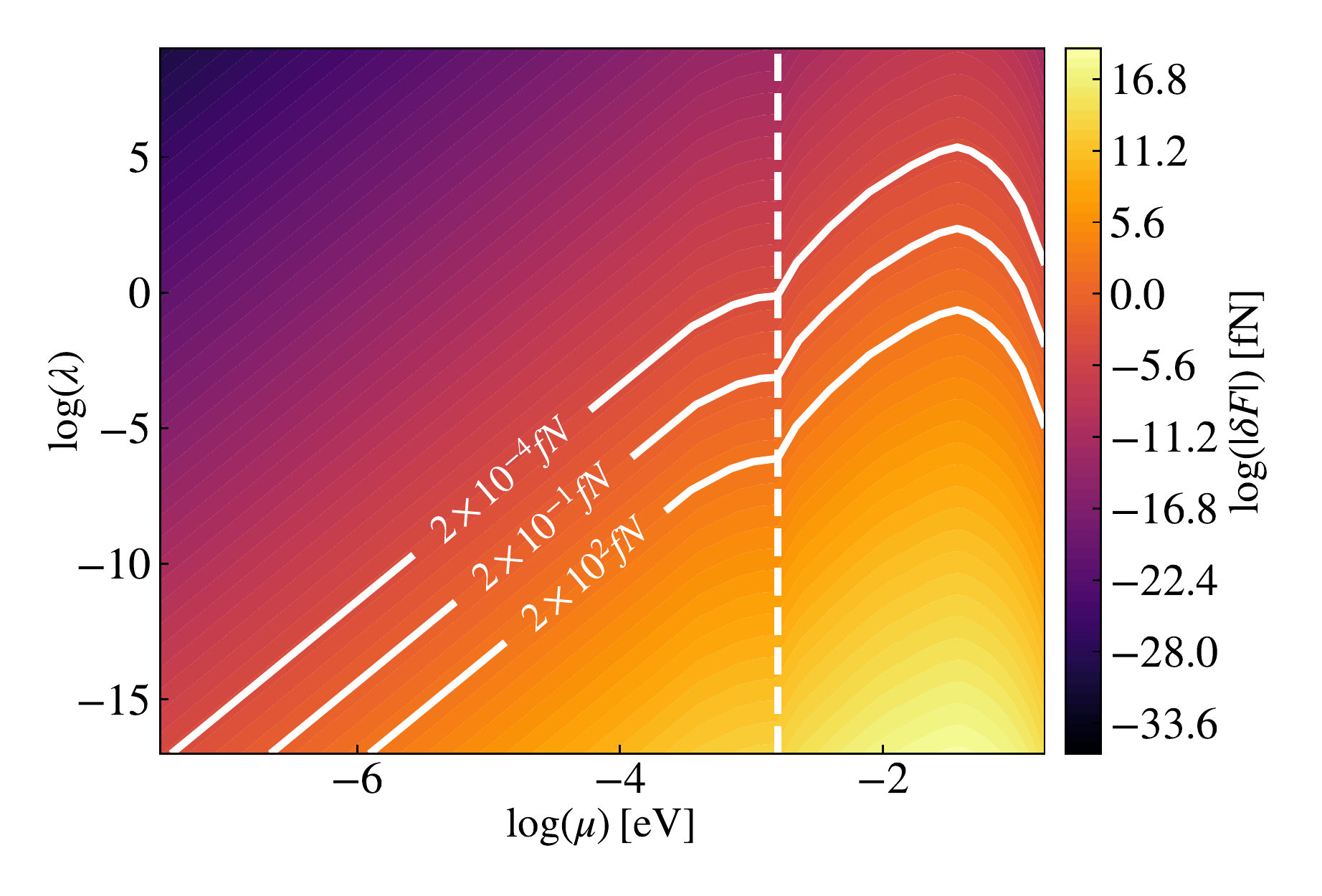}
\caption{\small \textbf{Left panel:} Ranges of $M$ and $\mu$ compatible with Eqs.~(\ref{eq:density_constraint}) and (\ref{eq:strong_screening}). The slanted lines are given by the condition of spontaneous symmetry breaking occurring in the vacuum and not inside the plate and the sphere. The horizontal upper bound is given by the requirement of the plate and the sphere being strongly screened. The parameter values outside the dark blue region do not satisfy at least one of Eqs.~(\ref{eq:density_constraint}) and (\ref{eq:strong_screening}), and therefore cannot be constrained by the present analysis. \textbf{Right panel:} Expected constraints from future plate-sphere based Casimir experiments on symmetron parameters $\mu$ and $\lambda$.  The experimental details are based on the current state of the art.  We have assumed that the experiment employs a large plate and a sphere of radius $R = 150 ~\mu$m, and measures the differential force between plate/sphere separations of $D = 15~\mu$m and $65~\mu$m. The solid, white curves correspond to three different upper bounds depending on the sensitivity of the experiment.
The vertical dashed line marks the value of $\mu$ for which the forces in the two configurations are almost identical, and the differential force signal is therefore negligibly small.  Note also that values of $\mu$ smaller than $R_\mathrm{vac}^{-1}$ are not constrainable.  We find that such a configuration is capable of constraining $\sim$ 5 orders of magnitude in $\mu$, which is remarkable given that laboratory experiments such as atom interferometry are currently sensitive to a window only $\sim$ 2 orders of magnitude wide.}
\label{fig:plate_sphere_constraints}
\end{figure*}

Having defined our window of sensitivity to $\mu$, we next consider the window of sensitivity to $M$. We have assumed that density of the sphere and plate is infinite, and the density of the surrounding gas is zero.  Quantitatively, for these assumptions to be valid the critical density of the symmetron $\rho_\mathrm{crit} = \mu^2 M^2$ must satisfy
\begin{equation}\label{eq:density_constraint}
\rho_\mathrm{vac} \ll \mu^2 M^2 \ll \rho_\mathrm{sphere}~.
\end{equation}
For our analysis throughout this paper we have taken $\rho_\mathrm{vac} = 10^2~\mathrm{eV}^4$ (corresponding to a pressure of $10^{-9}$ Torr for air at room temperature) and $\rho_\mathrm{sphere} = 10^{18}~\mathrm{eV}^4$ (corresponding to a density of $10~\text{g}~\text{cm}^{-3}$). Additionally, we have assumed that the sphere and the plate are strongly screened, so by Eq.~\eqref{screening-factor} we must restrict our attention to the regime
\begin{equation}\label{eq:strong_screening}
\frac{\rho_\mathrm{sphere} R^2}{M^2} \gg 1~.
\end{equation}
Let us mention that in this analysis the parameter $M$ does not have any effects on the value of the force. We have effectively discarded it by assuming $\rho_\mathrm{vac} = 0$ and $\phi = 0$ at the surfaces of the plate and the sphere. Therefore, the only way it can be constrained is through the two conditions we just presented. In the left panel of Fig.~\ref{fig:plate_sphere_constraints}, we show both of these constraints in the $\mu - M$ plane (note that $\lambda$ does not play any role in this discussion; see below). Particularly, the two slanted black lines correspond to the upper and lower limits on $M$ imposed through Eq.~(\ref{eq:density_constraint}), while the horizontal black line corresponds to the ratio in Eq.~(\ref{eq:strong_screening}) taken to be unity.

\begin{figure*}
\centering
\includegraphics[width = \textwidth]{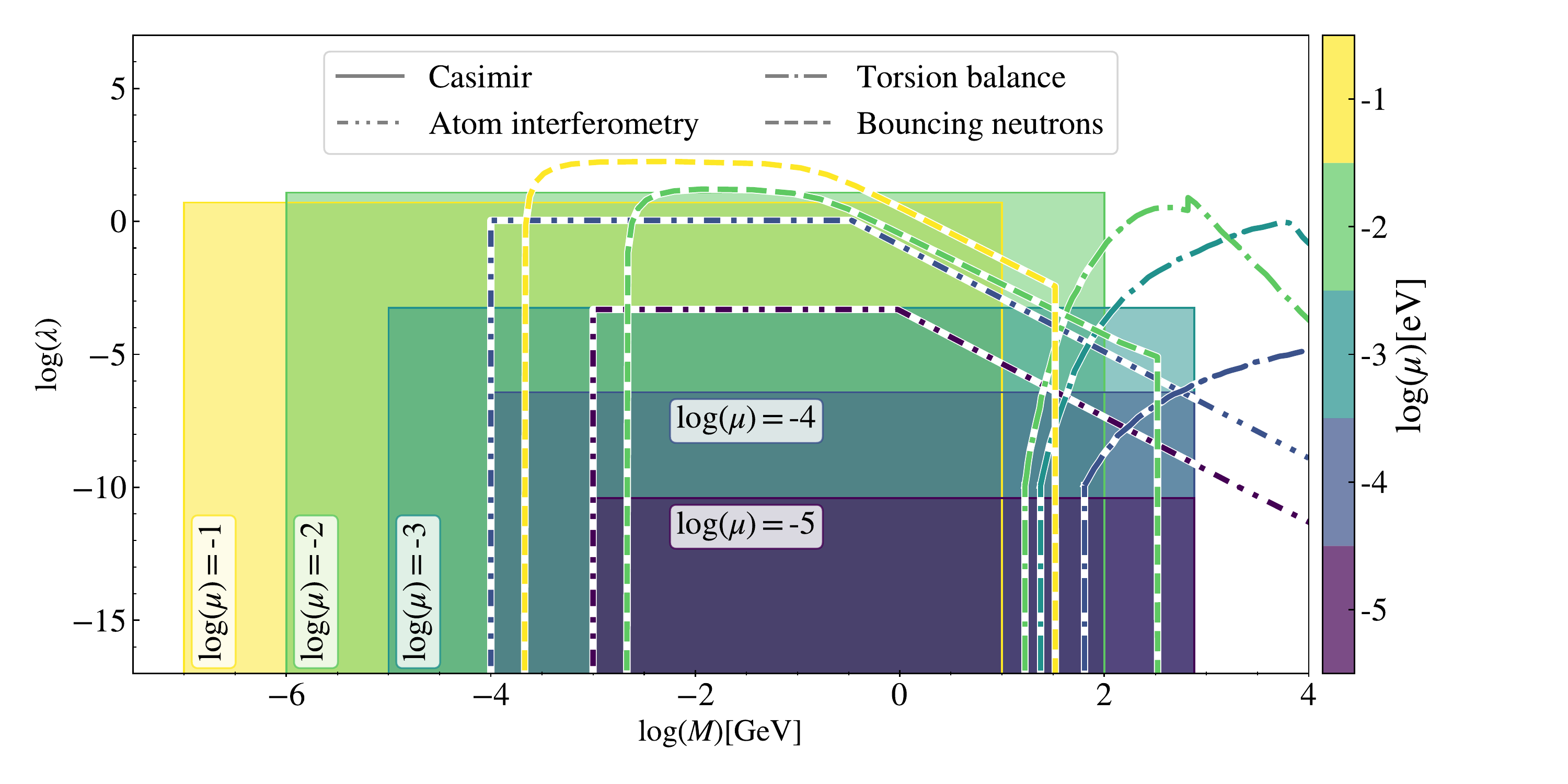}
\caption{\small Symmetron forecast constraints corresponding to a differential force upper bound of $\delta F_\mathrm{upper} = 0.2~ \mathrm{fN}$ at the $95\%$ confidence level, expected to be provided by future realistic Casimir experiments employing a sphere of radius $R = 150~\mu$m. Here we have computed the differential force between two configurations with $D_\mathrm{near} = 15 ~\mu$m and $D_\mathrm{far} = 65~\mu$m~.
The shaded regions will be excluded at a confidence level equal to the one provided by the Casimir experiment for the $\delta F$ measurement. Color indicates the symmetron mass $\mu$ to which a given constraint applies.  Also plotted are the relevant bounds from atom interferometry, torsion balance and ultracold bouncing neutron measurements reproduced from Refs.~\cite{Jaffe:2016fsh}, \cite{Upadhye:2012rc} and \cite{Cronenberg:2018qxf}, respectively.  It may be seen that (i) torsion balance experiments are complementary to Casimir experiments, and (ii) atom interferometry is extremely sensitive, but constrains only a relatively narrow window of $\mu$.  Bouncing neutrons result in strong constraints over a similar range of $\mu$, but presently apply only to a slightly narrower band in $M$.  This is largely due to the relatively large vacuum chamber density $\rho_\mathrm{vac}$ in bouncing neutrons experiments, which could plausibly be significantly improved in future experiments.
}
\label{fig:constraints}
\end{figure*}

We are now ready to present the constraints on the remaining two parameters $\mu$ and $\lambda$. While different values of $\mu$ correspond to different field profiles, and consequently different fifth forces, the effect of $\lambda$ on the force is much simpler. In fact, as we discussed earlier, it does not enter the numerical integration procedure. It simply enters only as a prefactor through Eq.~(\ref{eq:physical_force}), while $\hat{F}$ itself depends on $\mu$ (and not on $\lambda$). As a result, smaller values of $\lambda$ result in stronger fifth forces. Recall also that we have assumed $\lambda > 0$, which is required in order for the symmetron mechanism to work. Therefore, for particular choices of $\mu$ and $M$, we rule out the values of $\lambda$ that give differential forces $\delta F > \delta F_\mathrm{exp}$, where $\delta F_\mathrm{exp}$ is the bound reported by the Casimir experiment.

The forecast constraints in the $\lambda-\mu$ plane are shown in the right panel of Fig.~\ref{fig:plate_sphere_constraints} assuming different values of the experimental bound $\delta F_\mathrm{exp}$. The solid, white curves correspond to three different {\it hypothetical} upper bounds provided by future Casimir force experiments, consistent with the current state of the art. For this plot we have taken two configurations with $D/R = 15/150$ and $D/R = 65/150$, identical to the ones for the existing cutting-edge Casimir experiment described above. The vertical, dashed line marks the value of $\mu$ for which the forces in the two configurations are almost identical, and the differential force signal is therefore negligibly small.

For a given value of $\mu$, we can deduce the forecast constraints in the $\lambda-M$ plane. These constraints are presented in Fig.~\ref{fig:constraints}, alongside those currently provided by atom interferometry~\cite{Jaffe:2016fsh, Sabulsky:2018jma}, torsion balance~\cite{Upadhye:2012rc} and ultracold bouncing neutron experiments~\cite{Cronenberg:2018qxf}.  Note that the Casimir forecasts presented in this figure are obtained for the as-yet unperformed but realistic experiment described above.  We see that our forecasts are largely complementary to torsion balance constraints.  Atom interferometry provides very strong bounds, but only within a narrow window of $\sim 2$ orders of magnitude in $\mu$, while Casimir experiments are capable of constraining $\sim$ 5 orders of magnitude.

Bouncing neutrons provide powerful constraints over a similarly large window in $\mu$. This follows from the fact that the difference between energy levels of the bouncing neutrons in the gravitational field of the Earth vanishes only when $\mu z_0 \gg 1$, where $z_0=(2m^2 g)^{-1/3} \simeq 6$ microns is the typical size of the neutron's wave function over a perfect mirror.  As a result, those bounds \cite{Cronenberg:2018qxf} exist up to $\mu \sim$ eV, one order of magnitude larger than we presently forecast for the hypothetical but realistic Casimir experiment described above.  (It is possible that our estimates for $D_\mathrm{near}$ are slightly too conservative, in which case near-future Casimir experiments could also bound $\mu$ up to $\sim$ eV.)  It may also be seen from Fig.~\ref{fig:constraints} that bouncing neutrons constrain a slightly narrower band in $M$, mainly because of a relatively large vacuum chamber gas density $\rho_\mathrm{vac}$.  It is certainly plausible that future generations of the bouncing neutron experiment could remove this limitation.  Taken together, we find that near-future Casimir experiments will be competitive with, and complementary to, an assortment of existing bounds over a wide range of parameter space, chiefly thanks to the remarkably short distance scales probed.

\section{Conclusions}
\label{sec:conclusions}

In this paper we have presented solutions for the plate-sphere system in symmetron modified gravity.  We have identified two different analytical approximations that may be used when the sphere is much larger or smaller than the symmetron Compton wavelength, and have developed numerical solutions that naturally interpolate between these two regimes. Thanks to the scaling of the solutions with $\mu$, these solutions may find application across a wide range of scales, from microscopic laboratory to astrophysical setups.

To demonstrate the utility of our results we have applied them to Casimir force sensors, which often measure the force between a sphere and a plate.  Casimir force sensors have been previously unable to provide trustworthy bounds for symmetrons, largely due to the lack of detailed calculations predicting the symmetron force in such setups. Our results address this shortcoming. By considering a hypothetical Casimir experiment that is based on the current state of the art, we have found that Casimir force sensors hold a great deal of promise at probing a very large range of symmetron masses $\mu$ spanning $\sim 5$ orders of magnitude, thanks to the sensor's extraordinary sensitivity and to the short distance scales involved.

It will be interesting to apply our results to the actual Casimir experiments that will be coming in the near future.  Our results also pave the way for more detailed studies. For example it  may be necessary to model accurately the trench walls in Casimir experiments.  It is also interesting to apply an optimization technique to the experimental configuration parameters to find the values that maximize the sensitivity of the experiments to the symmetron force and the constraints on symmetron's parameters. We could also apply our results to astrophysical problems, such as configurations comprising a small spheroidal dwarf galaxy in the vicinity of a nearly planar spiral galaxy. Examples of such situations exist in the local group of galaxies. It would be extremely interesting to analyse the dynamical effects induced by the symmetron field on small, nearly spherical galaxies and determine if smaller values of the Compton wavelength $\mu^{-1}$ could be tested astrophysically. Notice that these effects would only be present if the symmetron does not couple to dark matter. On larger scales, one could also envisage testing the presence of symmetrons by studying the dynamics of galactic halos in the neighbourhood of galaxy filaments and great walls of galaxies. In this case, dark matter would be influenced by the presence of the symmetron. We leave these studies for future work.

\acknowledgements{
We are grateful to Clare Burrage, Jeremy Sakstein and Alessandra Silvestri for helpful discussions. B.E. is supported by a Leverhulme Trust Research Leadership Award. V.V. is supported by a de Sitter PhD fellowship of the Netherlands Organization for Scientific Research (NWO) and WPI Research Center Initiative, MEXT, Japan. Y.A. is supported by LabEx ENS-ICFP: ANR-10-LABX-0010/ANR-10-IDEX-0001-02 PSL*. A-C.D. acknowledges partial support from STFC under grants ST/L000385 and ST/L000636. This work is supported in part by the EU Horizon 2020 research and innovation programme under the Marie-Sklodowska grant No. 690575. This article is based upon work related to the COST Action CA15117 (CANTATA) supported by COST (European Cooperation in Science and Technology).  R.S.D. acknowledges support from the National Science Foundation through grants PHY-1607360 and PHY-1707985 and financial and technical support from the IUPUI Nanoscale Imaging Center, the IUPUI Integrated Nanosystems Development Institute, and the Indiana University Center for Space Symmetries. We wish to thank the Lorentz Center, Leiden, for their hospitality during the workshop `Dark Energy in the Laboratory' which led directly to this work.
}


\bibliography{symmetron}

\end{document}